\documentclass[graybox,secnum]{svmult}

\usepackage{amsfonts,amssymb,latexsym,amsmath,mathrsfs,amsbsy,bm}
\usepackage{slashed,bbold}
\usepackage{mathptmx}       
\usepackage{helvet}         
\usepackage{courier}        
\usepackage{type1cm}        
\usepackage{makeidx}         
\usepackage{graphicx}        
\usepackage{multicol}        
\usepackage[bottom]{footmisc}
\usepackage{hyperref}        
\usepackage{soul}            
\hypersetup{colorlinks=true,urlcolor=blue}
\usepackage[square,numbers]{natbib}
\makeindex             
\usepackage{amssymb}
\usepackage{amsmath}
\usepackage{amscd}
\usepackage{mdframed}
\begin{document}

\title*{Anomalies and the Green-Schwarz 
Mechanism\thanks{\href{https://doi.org/10.1007/978-981-19-3079-9_68-1}{Invited chapter}
for {\it Handbook of Quantum Gravity} 
(Eds. C.~Bambi, L.~Modesto, and I.~L.~Shapiro, Springer 2023).}}
\author{Luis \'Alvarez-Gaum\'e and Miguel \'A. V\'azquez-Mozo}
\institute{Luis \'Alvarez-Gaum\'e \at Simons Center for Geometry and Physics,
State University of New York Stony Brook,
NY-11794-3636, USA \& Theory Department CERN,
CH-1211 Geneva 23, Switzerland; \email{lalvarezgaume@scgp.stonybrook.edu}
\and Miguel \'A. V\'azquez-Mozo \at Departamento de F\'{\i}sica Fundamental, Universidad de Salamanca,
 Plaza de la Merced s/n,
 E-37008 Salamanca, Spain; \email{Miguel.Vazquez-Mozo@cern.ch}}
%
%
\maketitle

\abstract{Anomalies are a very powerful tool in constraining theories beyond the standard model.
We give a pedagogical overview of some topics illustrating the important role 
played by spacetime anomalies in string theory. After discussing the general problem of 
anomaly cancellation in quantum field theory, the focus is set on the cancellation of anomalies in
type-I string theory through the Green-Schwarz mechanism. 
The notion of anomaly inflow is also reviewed, as well as its application to the evaluation of 
D-brane anomalous couplings. 
Finally, we briefly comment on recent developments concerning 
the reformulation of anomalies in the language of category theory.
}

\keywords{Anomalies in quantum field theory and string theory; anomaly cancellation;
Green-Schwarz mechanism; anomaly inflow;
D-brane anomalous couplings.}

\section{The trouble with anomalies}

To the best of our knowledge, all fundamental interactions in nature are carried by 
quantum fields of spin one and two. Unlike scalar or spinors, the quantization of these fields 
requires special care in order to preserve locality and Lorentz invariance. 
A way to achieve this
is by introducing redundant degrees of freedom living in an extended 
Hilbert space~$\mathcal{H}$, so physical states are represented 
by equivalence classes under the action
of a group~$\mathcal{G}$. The freedom to switch from
one representative to another without changing the physics is 
what we call gauge invariance, and the space of physical states is obtained as
the quotient~$\mathcal{H}_{\rm phys}\equiv\mathcal{H}/\mathcal{G}$~\cite{QFT_book}. Unlike
standard physical symmetries mapping one state of the physical
Hilbert space into another, gauge invariance only changes the state representative, the label so to speak, 
acting thus trivially on~$\mathcal{H}_{\rm phys}$.

For this to work it is crucial to preserve the invariance under the choice of representative. Otherwise, spurious states in~$\mathcal{H}$ would enter~$\mathcal{H}_{\rm phys}$
rendering the theory nonunitary\footnote{The breakdown of gauge invariance also spoils 
renormalizability. From a modern effective field theory viewpoint, however, nonrenormalizability is 
no longer the dealbreaker it used to be.}. This may occur in chiral theories of the type of the standard model (SM), 
where gauge invariance can 
be broken in the process of quantization due to the necessary regularization.
Whenever this happens, the theory is said to be anomalous\footnote{Anomalies can also 
affect global symmetries whose breaking do not pose any threat to the theory's consistency. Here
we will not deal with these harmless anomalies.}~\cite{QFT_book,anomalies_rev,harvey_review}. 
The requirement
of chirality implies that gauge anomalies can only arise when the spacetime dimension is even.

Characterizing anomalies in gauge theories is thus of the utmost importance and its mandatory cancellation
poses very strict constraints on both the theory's spectrum and interactions.
The SM is a glaring example of this: the
condition that all gauge and mixed gauge-gravitational anomalies cancel leads to 
an essentially complete determination
of hypercharges, up to a global normalization. In the minimal supersymmetric
standard model (MSSM), on the other hand, the introduction of an additional Higgs doublet is necessary in order to cancel 
the anomaly induced by the chiral higgsino (see pages~\pageref{pag:SManomcan}-\pageref{pag:higgsino}).

At a diagrammatic level, anomalies are signaled by violations of the gauge Ward identities
induced by the parity-violating part of one-loop diagrams with gauge current insertions and 
chiral fields running in the loop.
In the case of the SM or their
supersymmetric extensions the relevant fields are the chiral fermions but, as we will see later, higher-dimensional 
theories also contain bosonic fields contributing to the anomaly.
On general grounds, in~$D=2n-2$ spacetime dimensions the anomaly is determined by 
a one-loop diagram with~$n$ current insertions, 
which in four dimensions gives the celebrated triangle diagram. One-loop diagrams
with more than~$n$ currents also contribute to the anomaly in non-Abelian theories, 
but their values are fully determined by 
the~$n$-point diagram, so one only needs to cancel the latter.

Functional methods offer a very 
powerful alternative to the diagrammatic analysis of anomalies. The basic object to consider
is the effective action~$\Gamma_{\rm eff}$,
obtained by integrating out all chiral degrees of freedom in the theory. Since these are massless 
fields,~$\Gamma_{\rm eff}$ is a nonlocal functional resumming all one-loop diagrams with an arbitrary
number of gauge current insertions.

Up to now, we have been talking about anomalies affecting any sort of gauge invariance, although they come in various kinds. 
It is however necessary to be more specific and distinguish among different types. First, we have 
gauge (consistent) anomalies, associated with the noninvariance of the effective action under 
infinitesimal gauge transformations~$\delta_{\chi}\mathcal{A}_{\mu}=\partial_{\mu}\chi+[\mathcal{A}_{\mu},\chi]$
\begin{align}
\delta_{\chi}\Gamma_{\rm eff}\neq 0.
\label{eq:gauge_anomaly_def1}
\end{align}
Their presence leads to quantum violations of the conservation of the corresponding gauge current.
For theories coupled to gravity, on the other hand, anomalies can also spoil 
the covariant conservation of the expectation value of the energy-momentum tensor,~$\nabla_{\mu}\langle T^{\mu\nu}\rangle \neq 0$.
These are called Einstein anomalies and are associated with the effective action's noninvariance
with respect to
infinitesimal diffeomorphisms
\begin{align}
\delta_{\rm diff}\Gamma_{\rm eff}\neq 0.
\label{eq:einstein_amomalies}
\end{align}
General relativity can also be understood as a gauge theory of local 
frame rotations,~$e^{a}_{\,\,\,\mu}\rightarrow \Lambda^{a}_{\,\,\,b}e^{b}_{\,\,\,\mu}$, 
where~$\Lambda^{a}_{\,\,\,b}$ is a Lorentz transformations and~$e^{a}_{\,\,\,\mu}$ is
the vielbein. This invariance can also be anomalous if gravity couples to 
chiral fields, giving rise to so-called 
Lorentz anomalies. They are signalled by a nonzero variation of the effective action under 
infinitesimal local Lorentz transformations~$\Lambda=\mathbb{1}+\epsilon$
\begin{align}
\delta_{\epsilon}\Gamma_{\rm eff}\neq 0.
\end{align}
In fact, it is always possible to find a local counterterm that added to
the effective action shifts
Lorentz into Einstein anomalies and vice versa~\cite{BZ_NPB,AG_GinspargAP}. 
We will take this into account and focus our attention on Lorentz anomalies from now on. 

In this chapter of the Handbook
we do not enter into a detailed discussion of quantum field theory anomalies, a subject that is
surveyed in a number of books and reviews (see, for instance,~\cite{anomalies_rev,harvey_review}). Instead, we 
limit our general presentation of anomalies to the basic recipes necessary to address the problem of their cancellation, with a focus
on the role played by this requirement in string theory. To this aim, two particular topics are 
selected. First, we carry out a detailed analysis of the workings of the Green-Schwarz (GS) mechanism, which was of pivotal importance in the 
development of string theory and goes well beyond the ten-dimensional open superstring theory in which 
it was first uncovered. Next, we study the seminal notion of anomaly inflow. 
To illustrate 
its implementation, we review how anomalous D-brane (and orientifold) couplings are determined by the condition that
worldvolume anomalies are cancelled by charge accretion/depletion. 
Our overview will be closed by a very brief discussion on the modern understanding of 
anomalies, that has led to a suggestive connection with category theory.

\section{Anomaly polynomials and Chern-Simons forms} 

\label{page:IRpole}
The main advantage of the effective action approach is that the anomaly can be computed 
from very general geometrical and topological considerations. It might be puzzling that this is possible at all,
given the fact that anomalies are usually seen as the result of UV ambiguities in the calculation of the 
Ward-Takahashi identities. This being true,
anomalies can also be interpreted as stemming from IR effects, signaled by zero-momentum poles in the expectation
value of the {\em anomalous currents themselves}. It is this IR side of anomalies that makes it
possible that they can be captured by studying the topological properties of the
vector bundles associated with the gauge and gravitational theories~\cite{nakahara}.

As a general rule, anomalies on a~$D$-dimensional curved manifold are determined
by an anomaly polynomial in $D+2=2n$ dimensions~\cite{zumino}, \label{page:anomaly_polynomial}
constructed from traces of powers of the spacetime 
curvature and the gauge 
field strength\footnote{In what follows, unless indicated otherwise, we use the
conventions of
ref.~\cite{nakahara}. Gauge fields are expressed in the language of differential 
forms~$\mathcal{A}=\mathcal{A}_{\mu}dx^{\mu}$ and, to avoid cluttering expressions, 
we drop the wedge sign~$\wedge$ in
exterior products whenever there is no risk of ambiguity.} 
\begin{align}
I_{2n}=P({\rm Tr\,}\mathcal{R}^{2},\ldots,{\rm Tr\,}\mathcal{R}^{n};{\rm tr\,}\mathcal{F},\ldots,
{\rm tr\,}\mathcal{F}^{n}).
\label{eq:anompoly_generic_form}
\end{align}
Here~$\mathcal{R}$ denotes 
the curvature 
two-form~$\mathcal{R}^{a}_{\,\,\,b}=d\omega^{a}_{\,\,\,b}+\omega^{a}_{\,\,\,c}\omega^{c}_{\,\,\,b}$
(with~$\omega^{a}_{\,\,b}$ the spin connection)
and~$\mathcal{F}$ is the gauge field strength, defined from the 
gauge potential~$\mathcal{A}$ by~$\mathcal{F}=d\mathcal{A}+\mathcal{A}^{2}$. 
In addition,~``${\rm Tr\,}$'' and~``${\rm tr\,}$'' respectively represent
trace over Lorentz and gauge indices in the appropriate representation. 
Expression~\eqref{eq:anompoly_generic_form} 
shows that the 
anomaly polynomial is invariant under local frame rotations and gauge transformations
\begin{align}
\mathcal{R}^{a}_{\,\,\,b}&\longrightarrow \Lambda^{a}_{\,\,\,\,c}
\mathcal{R}^{c}_{\,\,\,d}(\Lambda^{T})^{d}_{\,\,\,\,b}, \nonumber \\
\mathcal{F}&\longrightarrow g^{-1}\mathcal{F}g,
\end{align}  
with~$\Lambda^{a}_{\,\,\,\,b}\in\mbox{SO(1,$D-1$)}$ and~$g\in\mathcal{G}$, the Lorentz and gauge groups respectively.
By the index theorem, the integrated anomaly polynomial equals the 
index of certain differential operator in $D+2$ dimensions~\cite{AG_GinspargNPB,AG_GinspargAP,AG_Witten}.

Equation~\eqref{eq:anompoly_generic_form} 
also shows that the anomaly polynomial~$I_{2n}$ 
is a closed differential form,~$dI_{2n}=0$. In fact, by Poincar\'e's lemma, it is also locally
exact
\begin{align}
I_{2n}=dI^{0}_{2n-1},
\label{eq:CS_definition}
\end{align}
where~$I^{0}_{2n-1}$ is the Chern-Simons form. This~$(2n-1)$-form
can be integrated over a $(2n-1)$-dimensional 
manifold~$\mathcal{M}_{2n-1}$, whose boundary~$\partial\mathcal{M}_{2n-1}$ is identified with the
physical (Euclidean) spacetime. The result gives all the terms of the
one-loop quantum effective action associated with the anomaly
\begin{align}
\Gamma_{\rm eff}=2\pi i\int_{\mathcal{M}_{2n-1}}I^{0}_{2n-1}.
\label{eq:effectiveCS_action}
\end{align}
The global normalization of the integral is determined by either the index theorem or the diagrammatic
calculation.
From the point of view of the~$(2n-2)$-dimensional physical spacetime the action~\eqref{eq:effectiveCS_action} is nonlocal. 
This is to be expected, since~$\Gamma_{\rm eff}$ results from integrating out
a number of massless chiral fields.
Moreover, the Chern-Simons form is not
uniquely determined by eq.~\eqref{eq:CS_definition} since we are at liberty of 
shifting it by an arbitrary $(2n-2)$-form,~$I^{0}_{2n-1}\rightarrow I^{0}_{2n-1}+d\eta_{2n-2}$, without modifying it. 
This ambiguity reflects the freedom to 
add any local counterterm to the quantum effective action. Indeed, once integrated over~$\mathcal{M}_{2n-1}$ 
the previous shift amounts to
\begin{align}
\Gamma_{\rm eff}\longrightarrow \Gamma_{\rm eff}+2\pi i\int_{\partial\mathcal{M}_{2n-1}}\eta_{2n-2},
\end{align} 
where the last term, being an integral over physical space, is indeed local.

A crucial property of the Chern-Simons form is that,
unlike the anomaly polynomial, it is not gauge or Lorentz invariant. 
It can be shown, however, that its gauge variation is a
closed differential form,~$d\delta I^{0}_{2n-1}=0$, and therefore is also 
locally exact
\begin{align}
\delta I^{0}_{2n-1}=d I^{1}_{2n-2}.
\label{eq:descend_eqs}
\end{align}
Since we are taking infinitesimal transformations, the right-hand side is linear
in the gauge functions, as indicated by the superscript of~$I^{1}_{2n-2}$.
The anomaly is obtained by 
performing a gauge variation on the effective action~\eqref{eq:effectiveCS_action}
\begin{align}
\delta\Gamma_{\rm eff}=2\pi i\int_{\partial\mathcal{M}_{2n-1}}I^{1}_{2n-2}.
\label{eq:integrated_anomaly}
\end{align}
To be precise, this gives the so-called consistent anomaly. The name
implies that this form of the anomaly satisfies
the Wess-Zumino consistency conditions, a consequence of the fact that the commutator of two
gauge (resp. Lorentz) transformations acting on the effective action equals the action of the
transformation associated with their commutator.
Although the consistent anomaly is not generically gauge covariant, a 
covariant form of the anomaly can be obtained \label{pag:BZ_term}
by redefining the gauge current and adding counterterms~\cite{BZ_NPB}.

\section{Anomaly cancellation}

The previous discussion sets the framework to analyze anomaly cancellation: for a given theory,
we should find
its anomaly polynomial and see whether 
it is zero or not. Fortunately, there are powerful mathematical theorems at our disposal leading to
a direct computation of the anomaly polynomial for any kind of fields (see~\cite{anomalies_rev,nakahara}).
In the case of a Weyl fermion, the Atiyah-Singer index theorem in~$2n=D+2$ dimensions 
leads to the explicit expression
\begin{align}
I_{1\over 2}
=\big[\widehat{A}(\mathcal{M}){\rm ch}(\mathcal{F})\big]_{2n}.
\label{eq:index_spin1/2}
\end{align}
The subscript on the right-hand side indicates that we retain only the $2n$-form 
piece of the polynomial inside the brackets, and negative chirality spinors contribute with the
same polynomial and the opposite sign.
The term~$\widehat{A}(\mathcal{M})$ is 
called the $A$-roof (or Dirac) genus and is
determined by the curvature two-form. To give 
its explicit expression, we notice that~$\mathcal{R}^{a}_{\,\,\,b}$
defines a $(2n)\times(2n)$ antisymmetric
matrix of two-forms that can be 
brought to a block skew-diagonal form by an appropriate orthogonal transformation
\begin{align}
{1\over 2\pi}\mathcal{R}=\left(
\begin{array}{ccccc}
0 & x_{1} &  &  &  \\
-x_{1} & 0 &  &  & \\
 &  & 0 & x_{2} & \\
 &  & -x_{2} & 0  & \\
 &  &   &  & \ddots
\end{array}
\right).
\label{eq:curvature_matrix}
\end{align}
It is convenient to encode these skew eigenvalues in terms of the total Pontrjagin class
\begin{align}
p(\mathcal{R})=\prod_{i=1}^{n}(1+x_{i}^{2})\equiv 1+p_{1}(\mathcal{R})+p_{2}(\mathcal{R})+\ldots
+p_{n}(\mathcal{R}),
\end{align}
where the~$\ell$-th Poitrjagin class is 
defined as the homogeneous 
polynomial~\cite{anomalies_rev,nakahara}
\begin{align}
p_{\ell}(\mathcal{R})=\sum_{i_{1}<\ldots<i_{\ell}}^{n}x_{i_{1}}^{2}\ldots x_{i_{\ell}}^{2}.
\end{align}
These are~$4\ell$-forms. Noticing that
\begin{align}
{\rm Tr\,}\mathcal{R}^{2\ell}=(-1)^{\ell}\sum_{a=1}^{n}x_{a}^{2\ell}, \hspace*{1cm} 
{\rm Tr\,}\mathcal{R}^{2\ell+1}=0,
\end{align}
it is easy to write~$p_{\ell}(\mathcal{R})$ in terms of traces of powers of the curvature two-form. 
For the first few cases to be used later, we have
\begin{align}
p_{1}&=-{1\over 8\pi^{2}}{\rm Tr\,}\mathcal{R}^{2}, \nonumber \\
p_{2}&={1\over 128\pi^{4}}\Big[\big({\rm Tr\,}\mathcal{R}^{2}\big)^{2}-2{\rm Tr\,}\mathcal{R}^{4}\Big], \\
p_{3}&=-{1\over 3072\pi^{6}}\Big[\big({\rm Tr\,}\mathcal{R}^{2}\big)^{3}
-6\big({\rm Tr\,}\mathcal{R}^{2}\big)\big({\rm Tr\,}\mathcal{R}^{4}\big)+8{\rm Tr\,}\mathcal{R}^{6}\Big],
\nonumber
\end{align}
where, 
to simplify expressions, we omit the dependence of~$p_{k}$ on~$\mathcal{R}$.
Using Pontrjagin classes, the $A$-roof genus in~\eqref{eq:index_spin1/2} 
can be written as the polynomial 
\begin{align}
\widehat{A}(\mathcal{M})&\equiv 
\prod_{a=1}^{n}{x_{a}/2\over \sinh{(x_{a}/2)}} \label{eq:A-roof} \\
&=1-{1\over 24}p_{1}+{1\over 5760}\big(7p_{1}^{2}-4p_{2}\big)-{1\over 967680}\big(31p_{1}^{3}-44p_{1}p_{2}+16p_{3}\big)+\ldots
\nonumber
\end{align} 
As we will soon see, Pontrjagin classes are also useful in writing the contribution of other
fields to the gravitational anomaly.

The second factor on the right-hand side
of eq.~\eqref{eq:index_spin1/2}, containing all the dependence on the 
gauge field, is the Chern character. It is defined by the formal series
\begin{align}
{\rm ch}(\mathcal{F})&=\sum_{j=0}^{n}{1\over j!}{\rm tr\,}\left({i\over 2\pi}\mathcal{F}\right)^{j} \nonumber \\
&={\rm ch}_{0}(\mathcal{F})
+{\rm ch}_{1}(\mathcal{F})+\ldots+{\rm ch}_{n}(\mathcal{F}),
\label{eq:chern_character}
\end{align}
with the $2j$-form
\begin{align}
{\rm ch}_{j}(\mathcal{F})\equiv{1\over j!}{\rm tr}_{R}\left({i\over 2\pi}\mathcal{F}\right)^{j},
\end{align} 
defining
the $j$-th Chern character. The subscript~$R$ indicates that the traces are taken 
in the representation~$R$ in which the 
fermions transform, 
with~${\rm ch}_{0}(\mathcal{F})=N$ its dimension. 
Taking the exterior product of 
eqs.~\eqref{eq:A-roof} and~\eqref{eq:chern_character} and retaining the $2n$-form piece, we get the 
contribution of a chiral fermion to the anomaly polynomial in any even dimension. 

Both Pontrjagin classes and Chern characters are closed forms that can be written  
locally in terms of the corresponding gravitational and gauge Chern-Simons
forms as\footnote{To avoid dragging numerical factors around, our
definitions of the gauge and gravitational Chern-Simons forms~$\omega^{0}_{2n-1}$ and~$\Omega^{0}_{2n-1}$ 
do include the global normalizations of the traces in the
Chern characters and Pontrjagin classes. This differs from more usual conventions in the literature
which define~$d\omega^{0}_{2n-1}\equiv{\rm tr\,}\mathcal{F}^{2n}$.} 
\begin{align}
p_{\ell}(\mathcal{R})=d\Omega^{0}_{4\ell-1}(\mathcal{R},\omega), \hspace*{1cm}
{\rm ch}_{k}(\mathcal{F})=d\omega^{0}_{4k-1}(\mathcal{F},\mathcal{A}).
\label{eq:chern-simons_def1}
\end{align}
As explicitly indicated, 
these forms depend on the
spin connection and the gauge field respectively. Unlike their parent polynomials,
they are not invariant under local Lorentz
and gauge transformations. Their infinitesimal variations define differential 
forms through the descent equations
\begin{align}
\delta_{\epsilon}\Omega^{0}_{4\ell-1}(\mathcal{R},\omega)&=d\Omega^{1}_{4\ell-2}(\epsilon,\mathcal{R},\omega), \nonumber \\[0.2cm]
\delta_{\chi}\omega^{0}_{2k-1}(\mathcal{F},\mathcal{A})&=d\omega^{1}_{2k-2}(\chi,\mathcal{F},\mathcal{A}).
\label{eq:descent_1}
\end{align}
Further relations of this kind can be derived by considering the transformations 
of~$\Omega^{1}_{2\ell-2}$ and~$\omega^{1}_{2n-2}$, but we will not need them here. 
Notice that the expressions on the right-hand side are linear in the gauge functions~$\epsilon$ and~$\chi$, as it
is explicitly indicated by the superscripts.

After all these mathematical preliminaries, 
we are ready to analyze gauge and gravitational anomalies in theories where these are
brought about by Weyl fermions alone. 
To do so, we just need to account for all chiral fermions, add their contribution to the anomaly polynomial taking into account 
their chiralities and representations, and check whether the result vanishes.

\begin{svgraybox}
\centerline{\bf Example I: anomaly cancellation in the SM}\\

As an illustration, let us 
study the case of the SM where all potential anomalies are sourced by
spin-${1\over 2}$ fermions in the fundamental representation of the gauge group. 
Using eq.~\eqref{eq:index_spin1/2}, together with 
the expansions~\eqref{eq:A-roof} and~\eqref{eq:chern_character}, we can write the 
six-form anomaly polynomial relevant to four dimensions:\label{pag:SManomcan}
\begin{align}
I_{1\over 2}[\mathcal{F},\mathcal{R}]&=
{\rm ch}_{3}-{1\over 24}p_{1}{\rm ch}_{1}=-{i\over 48\pi^{3}}\left[{\rm tr}_{\rm f}\mathcal{F}^{3}
-{1\over 8}\big({\rm Tr\,}\mathcal{R}^{2}\big)\big({\rm tr}_{\rm f}\mathcal{F}\big)
\right],
\label{eq:anomaly_D=4}
\end{align}
where, for simplicity, we dropped the dependence on~$\mathcal{F}$ in the Chern characters
and we indicated that all traces
are evaluated in the fundamental representation of~$\mbox{SU(2)}_{L}\times\mbox{U(1)}_{Y}$. They in fact
can be easily computed in terms of traces over
the corresponding group factors
\begin{align}
{\rm tr}_{\rm f}\mathcal{F}^{3}&=2{\rm tr}_{\rm f}\mathcal{F}_{Y}^{3}+3\big({\rm tr}_{\rm f}\mathcal{F}_{Y}\big)
\big({\rm tr}_{\rm f}\mathcal{F}_{L}^{2}\big)+{\rm tr}_{\rm f}\mathcal{F}_{L}^{3}, \nonumber \\
{\rm tr}_{\rm f}\mathcal{F}&=2{\rm tr}_{\rm f}\mathcal{F}_{Y}.
\label{eq:tracesF_D=4}
\end{align}
where~$\mathcal{F}_{L}$ and~$\mathcal{F}_{Y}$ denote the SU(2)$_{L}$ and U(1)$_{Y}$ field 
strengths, respectively, and we have used that~${\rm tr}_{\rm f}\mathcal{F}_{L}=0$ due to the tracelessness of the
generators of~SU(2)$_{L}$. 
We have to include in the anomaly polynomial the contributions of all Weyl 
fermions, weighted with the signs corresponding to their helicities. 
Keeping in mind that SU(2)$_{L}$ does not couple to positive-chirality
fermions, we have:
\begin{align}
\sum_{+}{\rm tr}_{\rm f}\mathcal{F}^{3}_{+}-\sum_{-}{\rm tr}_{\rm f}\mathcal{F}^{3}_{-}
&=2\left(\sum_{+}{\rm tr}_{\rm f}\mathcal{F}^{3}_{Y,+}-\sum_{-}{\rm tr}_{\rm f}\mathcal{F}^{3}_{Y,-}\right)
\nonumber \\
&-3\sum_{-}\big({\rm tr}_{\rm f}\mathcal{F}_{Y,-}\big)\big({\rm tr}_{\rm f}\mathcal{F}_{L,-}^{2}\big)
-\sum_{-}{\rm tr}_{\rm f}\mathcal{F}^{3}_{L,-}.
\end{align}
The last term on the right-hand side of this expression vanishes because of the 
identity~${\rm tr}_{\rm f}\mathcal{F}^{3}_{L,-}\sim{\rm tr\,}\big(\sigma^{(i}\sigma^{j}\sigma^{k)}\big)=0$. As for the second one, 
all left-handed fields couple with the same strength to the~SU(2)$_{L}$ gauge bosons 
so~${\rm tr}_{\rm f}\mathcal{F}_{L,-}^{2}$ factors out of the sum. Summing
over all leptons and (three-colored) quarks within a single family,
we find the relevant traces to vanish 
\begin{align}
\sum_{+}{\rm tr}_{\rm f}\mathcal{F}^{3}_{Y,+}-\sum_{-}{\rm tr}_{\rm f}\mathcal{F}^{3}_{Y,-}&\sim
\sum_{+}Y_{+}^{3}-\sum_{-}Y_{-}^{3}
=3\times 2\times \left({1\over 6}\right)^{3}+2\times\left(-{1\over 2}\right)^{3} \nonumber \\
&-3\times\left({2\over 3}\right)^{3}
-3\times\left(-{1\over 3}\right)^{3}-(-1)^{2}=0, \\
\sum_{-}{\rm tr}_{\rm f}\mathcal{F}_{Y,-}&\sim \sum_{-}Y_{-}=3\times 2\times\left({1\over 6}\right)
+2\times\left(-{1\over 2}\right)=0.
\nonumber
\end{align}
This ensures that pure gauge anomalies cancel family by family. We are left just with the mixed 
gauge-gravitational term on the right-hand side of~\eqref{eq:anomaly_D=4}.
Since in four dimensions 
gravity treats both chiralities in the same way,
we only need to use the second line of~\eqref{eq:tracesF_D=4} 
to find 
\begin{align}
\sum_{+}{\rm tr}_{\rm f}\mathcal{F}_{+}-\sum_{-}{\rm tr}_{\rm f}\mathcal{F}_{-}&\sim
\sum_{+}Y_{+}-\sum_{-}Y_{-}
=3\times 2\times\left({1\over 6}\right)+2\times\left(-{1\over 2}\right) \nonumber \\
&-3\times\left({2\over 3}\right)
-3\times\left(-{1\over 3}\right)-(-1)=0.
\end{align}
Thus, mixed gauge-gravitational anomalies also cancel within a single family. 
The SM is therefore anomaly-free.

Anomaly cancellation in the SM completely fixes the hypercharges up to a global normalization.
An interesting exercise is to consider the gauge group to 
be~$\mbox{SU}(N_{c})\times\mbox{SU}(2)\times U(1)$ with the same family structure as in the SM
but leaving the hypercharges of the fermion fields 
undetermined. The anomaly cancellation conditions become then a set of 
homogeneous equations in the hypercharges which, normalizing the hypercharge of the right-handed 
electron~as~$Y(e_{R})=-1$, has a unique solution. An unphysical solution also exists 
when~$Y(e_{R})=0$.

The different 
diagrammatic contributions to the gauge anomaly  can easily read from
eqs.~\eqref{eq:anomaly_D=4} and~\eqref{eq:tracesF_D=4}. The three terms in the first line 
of the latter equation correspond
to triangle diagrams with three hypercharge, one hypercharge and two~SU(2)$_{L}$, and
three SU(2)$_{L}$ currents, respectively. The second line, once multiplied by~${\rm Tr\,}\mathcal{R}^{2}$, gives
the contribution of a triangle
with one hypercharge and two graviton insertion coupling to the fermion 
via the energy-momentum tensor (see, for example,~\cite{QFT_book} for more details).

As a bonus, we can work out the conditions for anomaly cancellation in the MSSM almost for free. 
Besides the chiral fields of the SM, whose anomalies we have seen are canceled family by family, 
its minimal supersymmetric extension only includes two
kinds of potentially dangerous fields: the gauginos and the higgsino. The first kind, transforming
in the adjoint
representation of the SM gauge group, do not give any nonzero contribution to the anomaly polynomial. As for
the higgsino, it is an SU(2)$_{L}$-doublet Weyl fermion with hypercharge~$1\over 2$. 
Its inclusion results in the nonvanishing traces\label{pag:higgsino}
\begin{align}
{\rm tr\,}\mathcal{F}_{Y}^{3}\sim \left({1\over 2}\right)^{3}, 
\hspace*{1cm} {\rm tr\,}\mathcal{F}_{Y}\sim {1\over 2}.
\end{align} 
The upshot to this is that the MSSM with a single Higgs field suffers from gauge and mixed anomalies.
To fix the situation, a second Higgs doublet has to be added to the one present in the SM, 
with an oposite chirality higgsino canceling the anomaly induced by the first one.~$\blacksquare$

\end{svgraybox}

As explained above, 
diffeomorphism or local Lorentz invariance can be anomalous if parity is broken~\cite{AG_Witten}. 
This only occurs
when the dimension of the spacetime satisfies~$D=4k+2$ (i.e., $n=2k+2$), with integer~$k$.
This includes the case $D=2$ and $D=10$, of 
particular interest to string theorists. When~$D=4k$~($n=2k+1$), as it is the case of 
our four-dimensional world, \label{pag:gravanomD=4k+2}
there are no pure gravitational anomalies\footnote{At a physical level, this
happens because in~$D=4k$ CPT reverses chirality. Thus, CPT-invariant theories contain 
as many left-handed as right-handed fermions. Since the equivalence principle states that gravity
couples to matter universally, one chirality necessarily cancels the contribution
to the gravitational anomaly of the opposite one. When~$D=4k+2$, 
however, CPT preserves chirality and a mismatch 
in the number of left- and right-handed fermons is allowed. Pure gravitational anomalies 
do not cancel automatically in this case.}. This is why we did not have to care about them in our analysis of anomaly
cancellation in the SM or the MSSM, although we did indeed pay attention to the 
gravitational contribution to the gauge anomaly.

As in the gauge case, the gravitational anomaly in~$D$ dimensions
is computed from a certain anomaly polynomial
in dimension~$D+2=2n$~\cite{AG_Witten}. For an uncharged Weyl fermion of positive
(resp. negative) helicity, 
it is given by
the same $A$-roof genus defined in eq.~\eqref{eq:A-roof}
\begin{align}
I_{1\over 2}[\mathcal{R}]=\pm[\widehat{A}(\mathcal{M})]_{2n}.
\label{eq:I1/2_grav_xi}
\end{align}
In supergravity (SUGRA) theories, a Weyl spin-${3\over 2}$ gravitino can also be the
source gravitational anomalies, although being neutral it does not contribute to the gauge anomaly. 
Its anomaly polynomial 
is given by the~$2n$-form piece of~\cite{anomalies_rev,AG_GinspargAP}
\begin{align}
I_{3\over 2}[\mathcal{R}]&\equiv\left[\prod_{a=1}^{n}{x_{a}/2\over \sinh{(x_{a}/2)}}\right]
\left(2\sum_{b=1}^{n}\cosh{x_{b}}-1\right) \nonumber \\[0.2cm]
&=2n-3+{27-2n\over 24}p_{1}+{1\over 5760}\Big[(219+14n)p_{1}^{2}-4(237+2n)p_{2}\Big]
\label{eq:I3/2_grav_xi} \\[0.2cm]
&+{1\over 967680}\Big[(597-62n)p_{1}^{3}-4(537-22n)p_{1}p_{2}+16(507-2n)p_{3}\Big]+\ldots
\nonumber
\end{align}
Notice that, unlike the $A$-roof genus,~$I_{3\over 2}$ explicitly depends on the dimension.

Our SM intuition might lead us to think that chiral fermions are the only fields producing 
gravitational anomalies and that we are done with our search of relevant polynomials to address
the problem of anomaly cancellation in any dimension. 
This is not the case. Bosonic
degrees of freedom can also play a role if they are a source of parity breaking. This 
is what happens if we have
an $r$-form field $B_{r}$ whose field strength is either self-dual~($+$) or anti-self-dual~($-$)
\begin{align}
dB_{r}=\pm \star dB_{r},
\end{align}
where the star indicates the Hodge dual and the condition can be satisfied only for~$r=(D-2)/2$. 
Since the star operator picks up a minus sign under parity, 
the (anti)self-dual condition is not parity invariant and these fields should be taken into account 
when analyzing anomalies. 
The contribution 
of a self-dual tensor field to the anomaly is given in terms of the Hirzebruch 
polynomial~$L(\mathcal{M})$ as
\begin{align}
I_{\rm sd}[\mathcal{R}]&=-{1\over 8}L(\mathcal{M})\equiv-{1\over 8}\prod_{a=1}^{n}{x_{a}\over\tanh{x_{a}}} 
\label{eq:Isd_grav} \\[0.2cm]
&=-{1\over 8}-{1\over 24}p_{1}+{1\over 360}\big(p_{1}^{2}-7p_{2}\big)
-{1\over 7560}\big(2p_{1}^{3}-13p_{1}p_{2}+62p_{3}\big)+\ldots,
\nonumber
\end{align}
while for an anti-self-dual tensor field the polynomial has the opposite sign. Since these fields are 
uncharged under the gauge group, there are no terms depending on the gauge field strength. 

A look at 
expressions~\eqref{eq:A-roof},~\eqref{eq:I3/2_grav_xi}, and~\eqref{eq:Isd_grav} shows that 
all their monomials 
are differential forms whose ranks are multiples of four. This means
they only contain terms of rank~$2n$ if~$n=2k$, or equivalently~$D=4k+2$. This reflects the 
fact stated in page~\pageref{pag:gravanomD=4k+2} that
pure gravitational anomalies are only possible in these dimensions.

\begin{svgraybox}
\centerline{\bf Example II: anomaly cancellation in type-IIB SUGRA}\\

\nopagebreak
As a second instance of anomaly cancellation, 
let us analyze~$\mathcal{N}=2$ ten-dimensional type-IIB 
SUGRA~\cite{AG_Witten}, the
low-energy limit of type-IIB closed string theory. This is a chiral
theory without gauge interactions where 
gravitational anomalies may arise. Its spectrum contains a number of potentially dangerous fields: 
two negative chirality spin-${3\over 2}$ gravitini, 
two positive chirality spin-${1\over 2}$ dilatinos, 
and a self-dual four-form field. All chiral fermions satisfy in addition the Majorana condition. 
In ten dimensions, the relevant anomaly polynomial is a 12-form that
in the case of the spin-${1\over 2}$ fields is read from eqs.~\eqref{eq:A-roof} and~\eqref{eq:I1/2_grav_xi}, with the result
\begin{align}
I_{1\over 2}[\mathcal{R}]&=-{1\over 967680}\big(31p_{1}^{3}-44p_{1}p_{2}+16p_{3}\big),
\label{eq:I1/2_grav}
\end{align}
whereas for the gravitino [cf.~\eqref{eq:I3/2_grav_xi}] we have
\begin{align}
I_{\rm 3\over 2}[\mathcal{R}]&={1\over 107520}\big(25p_{1}^{3}-180p_{1}p_{2}+880p_{3}\big).
\label{eq:I3/2_grav}
\end{align}
Finally, we have 
the contribution of the self-dual four-form field given in eq.~\eqref{eq:Isd_grav}
\begin{align}
I_{\rm sd}[\mathcal{R}]&=-{1\over 7560}\big(2p_{1}^{3}-13p_{1}p_{2}+62p_{3}\big).
\end{align}
The anomaly polynomial~$I_{12}$ is given by
\begin{align}
I_{\rm 12}=-2\times{1\over 2}I_{1\over 2}[\mathcal{R}]
+2\times{1\over 2}I_{3\over 2}[\mathcal{R}]+I_{\rm sd}[\mathcal{R}].
\end{align}
The factors of~2 in front of the gravitino and
dilatino contributions 
reflect the fact that there are two species of each kind, whereas the~${1\over 2}$'s are there because
these fermions satisfy the Majorana-Weyl condition, which halves the number of 
real degrees of freedom with respect to 
a Weyl fermion. Their sign, in turn, is determined by their respective chiralities. 
Using the explicit expressions given above for each term,
we check that
\begin{align}
-I_{1\over 2}[\mathcal{R}]+I_{3\over 2}[\mathcal{R}]+I_{\rm sd}[\mathcal{R}]=0.
\end{align}
Thus, all gravitational anomalies cancel in type-IIB SUGRA. Remember that this theory does 
not contain gauge fields, so we do not need to care about either gauge or mixed anomalies.~$\blacksquare$

\end{svgraybox}

In ten dimensions, besides~$\mathcal{N}=2$ type-IIB SUGRA,
there is another interesting chiral theory:~$\mathcal{N}=1$ SUGRA that, in addition to a graviton,
a dilaton, and a two-form field, also contains a left-handed gravitino and a right-handed dilatino, 
both of the Majorana-Weyl type. It was found in ref.~\cite{AG_Witten} that
this theory is not free from 
gravitational anomalies. Indeed, using eqs.~\eqref{eq:I1/2_grav} and~\eqref{eq:I3/2_grav}
we see the total anomaly polynomial is nonzero
\begin{align}
I_{12}&={1\over 2}I_{\rm 3\over 2}[\mathcal{R}]
-{1\over 2}I_{\rm 1\over 2}[\mathcal{R}] \nonumber \\
&={1\over 15120}\big(2p_{1}^{3}-13p_{1}p_{2}+62p_{3}\big)\neq 0,
\label{eq:I_12_N=1SUGRA}
\end{align}
where again the~${1\over 2}$ factors are due to the Majorana condition.  

This result was a source of concern given the relation of this theory 
to type-I superstrings, which  
in the 1980s were regarded as promising candidates for a unified
theory of all four interactions. Its massless spectrum contains, besides
a~$\mathcal{N}=1$ SUGRA multiplet, 
a~$\mathcal{N}=1$ super-Yang-Mills (SYM) multiplet including 
a gauge boson and its gaugino, a Majorana-Weyl left-handed fermion, 
both transforming in the adjoint representation of the gauge group. To check 
anomaly cancellation in type-I string theory, 
we add to eq.~\eqref{eq:I_12_N=1SUGRA} the contribution from the gaugino
\begin{align}
I_{12}={1\over 2}I_{\rm 3\over 2}[\mathcal{R}]
-{1\over 2}I_{\rm 1\over 2}[\mathcal{R}]
+{1\over 2}I_{1\over 2}[\mathcal{F},\mathcal{R}],
\end{align}
where last term is computed from~\eqref{eq:index_spin1/2}
\begin{align}
I_{1\over 2}[\mathcal{F},\mathcal{R}]&={\rm ch}_{6}-{1\over 24}p_{1}{\rm ch}_{4}+{1\over 5760}\big(7p_{1}^{2}-4p_{2}\big){\rm ch}_{2} \nonumber \\
&-{N\over 967680}\big(31p_{1}^{3}-44p_{1}p_{2}+16p_{3}\big),
\label{eq:I1/2_12}
\end{align}
with all Chern characters evaluated in the adjoint representation.
Putting all terms together, we get the anomaly polynomial
of type-I SUGRA
\begin{align}
2I_{12}&={\rm ch}_{6}-{1\over 24}p_{1}{\rm ch}_{4}+{1\over 5760}\big(7p_{1}^{2}-4p_{2}\big){\rm ch}_{2} 
\nonumber \\
&+{256-31N\over 967680}p_{1}^{3}+{11N-416\over 241920}p_{1}p_{2} +{496-N\over 60480}p_{3},
\label{eq:I_12_N_general}
\end{align}
where~$N$ is the dimension of the adjoint representation of the gauge group. 
This nonvanishing result seems to imply that 
type-I string theory is anomalous and therefore should be discarded. More precisely, inspecting the anomaly polynomial we verify 
the existence of gauge, gravitational, and mixed anomalies. The first two are associated 
with hexagon diagrams with six gauge fields and six graviton fields respectively. 
Mixed anomalies, on the other hand, arise from hexagon diagrams with four gauge fields and two gravitons 
and two gauge fields and four gravitons. \label{page:hexagon_10D}

\section{The Green-Schwarz solution}
\label{sec:GS}

Despite its bad prospects, in 1984 Michael Green and John Schwarz~\cite{GS1,GS2} 
showed that type-I string theory is anomaly-free for a particular choice of the
gauge group\footnote{The Green-Schwarz mechanism is discussed 
in most books and reviews on string theory.
See, for example,~\cite{string_reviews,johnson} and particularly~\cite{GSW}.}.
They uncovered the existence of a nontrivial mechanism 
to cancel all anomalies in the theory
involving the massless bosonic two-form 
field present
in the spectrum of~$\mathcal{N}=1$ SUGRA. In principle this might sound surprising, since
one would not expect this field to contribute
to the anomaly. Moreover, as we will see, the cancellation terms come from tree-level diagrams.

The GS mechanism
is based on the observation that the anomaly polynomial computed in~\eqref{eq:I_12_N_general} 
can be canceled provided it admits the factorization
\begin{align}
I_{12}=\big(\lambda {\rm ch}_{2}+p_{1}\big)X_{8},
\label{eq:factorizationGS}
\end{align}
with~$\lambda$ a constant and~$X_{8}$ an eight-form that locally can be written as~$X_{8}=dX_{7}^{0}$.
Let us discuss first how this fact leads to the cancellation of the anomaly, addressing later 
the problem of the conditions
required for the factorization~\eqref{eq:factorizationGS} to occur.
 
We know the anomaly polynomial is an exact form,~$I_{12}=dI_{11}^{0}$, and
it is possible to show that there is a solution for~$I_{11}^{0}$ given by
\begin{align}
I_{11}^{0}={1\over 2}\big(\lambda \omega_{3}^{0}+\Omega_{3}^{0}\big)X_{8}+
{1\over 2}\big(\lambda {\rm ch}_{2}+p_{1}\big)X_{7}^{0}
+{\alpha\over 2} d\big[\big(\lambda \omega_{3}^{0}+\Omega_{3}^{0}\big)
X_{7}^{0}\big],
\label{eq:I_11_first_factor}
\end{align}
where we introduced the gravitational and gauge Chern-Simons forms~$\Omega_{3}^{0}$ and~$\omega_{3}^{0}$ 
defined in~\eqref{eq:chern-simons_def1}. 
We also included the last exact term proportional to an arbitrary coefficient~$\alpha$, thus
exploiting the freedom of adding local counterterms to the quantum effective action. 
Performing a gauge transformation, we get
\begin{align}
\delta I^{0}_{11}&={1-\alpha\over 2}\big(\lambda d\omega_{2}^{1}+d\Omega_{2}^{1}\big)X_{8}
+{1+\alpha\over 2}\big(\lambda {\rm ch}_{2}+p_{1}\big)dX_{6}^{1},
\end{align}
after applying the descent relation~$\delta X_{7}^{0}=dX_{6}^{1}$.
Since both~$p_{2}$ and~${\rm ch}_{2}$ are closed forms, we see that locally~$\delta I^{0}_{11}=dI^{1}_{10}$
with
\begin{align}
I^{1}_{10}&={1-\alpha\over 2}\big(\lambda \omega_{2}^{1}+\Omega_{2}^{1}\big)X_{8}
+{1+\alpha\over 2}\big(\lambda \omega_{3}^{0}+\Omega_{3}^{0}\big)dX_{6}^{1}.
\end{align}
The anomalous variation of the quantum effective action is then obtained by
integrating over the ten-dimensional 
spacetime~[cf.~\eqref{eq:integrated_anomaly}]
\begin{align}
\delta\Gamma_{\rm eff}
={1-\alpha\over 2}c\int_{\mathcal{M}_{10}}\big(\lambda \omega_{2}^{1}+\Omega_{2}^{1}\big)X_{8}
+{1+\alpha\over 2}c\int_{\mathcal{M}_{10}}\big(\lambda \omega_{3}^{0}+\Omega_{3}^{0}\big)dX_{6}^{1}.
\label{eq:anomaly_general_lambda}
\end{align}
where~$c$ stands for the overall normalization of the action.

The question is whether this can be canceled by the variation 
of a local counterterm. It is in fact possible,
provided our theory contains a two-form field~$B$ with the following gauge transformation
\begin{align}
\delta C_{2}=\lambda \omega_{2}^{1}+\Omega_{2}^{1}.
\label{eq:C2_gauge_trans}
\end{align}
When this happens, it is straightforward to check that the local counterterm
\begin{align}
\Gamma_{\rm ct}=-c\int_{\mathcal{M}_{10}}C_{2} X_{8}
-{1+\alpha\over 2}c\int_{\mathcal{M}_{10}}\big(\lambda\omega^{0}_{3}+\Omega^{0}_{3}\big)X_{7}^{0}.
\label{eq:GS_counterterm}
\end{align}
has a gauge variation that exactly cancels the anomaly~\eqref{eq:anomaly_general_lambda}
\begin{align}
\delta\Gamma_{\rm ct}&={\alpha-1\over 2}c
\int_{\mathcal{M}_{10}}\big(\lambda\omega_{2}^{1}+\Omega_{2}^{1}\big)X_{8}
-{1+\alpha\over 2}c\int_{\mathcal{M}_{10}}\big(\lambda\omega_{3}^{0}+\Omega_{3}^{0}\big)dX_{6}^{1}.
\end{align}
At the same time, the new gauge variation of~$C_{2}$ implies that its field strength has to be defined as 
\begin{align}
H=dC_{2}-\lambda\omega_{3}^{0}-\Omega_{3}^{0},
\label{eq:HmodifiedGS}
\end{align} 
so it remains gauge invariant.

\paragraph{\bf Constraints on the gauge group.}

We have seen how the factorization~\eqref{eq:factorizationGS} leads to the cancellation of
all anomalies. What we still do not know, however, is whether
this factorization can be actually achieved.

A look at the anomaly polynomial~\eqref{eq:I_12_N_general} makes it clear that the obstruction to the
sought factorization
lies in the presence of the sixth Chern character,~$\mbox{ch}_{6}$, and the third Pontrjagin class,~$p_{3}$.
The latter term can be easily removed by restricting the gauge group to those whose 
adjoint representation has dimension~$N=496$. Doing so, the anomaly polynomial simplifies to
\begin{align}
2I_{12}&={\rm ch}_{6}-{1\over 24}p_{1}{\rm ch}_{4}+{1\over 5760}\big(7p_{1}^{2}-4p_{2}\big){\rm ch}_{2} 
-{1\over 64}p_{1}^{3}+{1\over 48}p_{1}p_{2}.
\label{eq:anomaly_typeI}
\end{align}
This leaves us with the problem of~$\mbox{ch}_{6}$. 
The only way to circumvent it
is by further restricting the gauge group to those 
satisfying a relation of the 
type~${\rm ch}_{6}=A{\rm ch}_{2}{\rm ch}_{4}+B{\rm ch}_{2}^{3}$, for some constants~$A$ and~$B$. 
Writing~$X_{8}$ as a linear combination 
of~${\rm ch}_{4}$,~${\rm ch}_{2}^{2}$,~$p_{1}{\rm ch}_{2}$,~$p_{1}^{2}$,
and~$p_{2}$ with undetermined coefficients and implementing the factorization~\eqref{eq:factorizationGS},
we find a unique solution where
\begin{align}
{\rm ch}_{6}={1\over 720}\left({\rm ch}_{2}{\rm ch}_{4}-{1\over 1800}{\rm ch}_{2}^{3}\right),
\label{eq:relation_chern_classes}
\end{align}
$\lambda=-{1\over 30}$, and~$X_{8}$ is given by 
\begin{align}
X_{8}&=-{1\over 48}\left(
{\rm ch}_{4}-{1\over 1800}{\rm ch}_{2}^{2}
-{1\over 60}p_{1}{\rm ch}_{2}
+{3\over 8}p_{1}^{2}-{1\over 2}p_{2}
\right).
\label{eq:X8_general10D}
\end{align}
In all previous expressions we have to keep in mind that all Chern characters are computed 
from traces in the adjoint representation of the gauge group.

The question remains as to whether there are any groups whose adjoint representations have
dimension~$N=496$ and at the same time satisfy the 
relation~\eqref{eq:relation_chern_classes}, which we recast as the following identity
among traces
\begin{align}
{\rm tr}_{\rm adj}\mathcal{F}^{6}
={1\over 48}\big({\rm tr}_{\rm adj}\mathcal{F}^{2}\big)
\big({\rm tr}_{\rm adj}\mathcal{F}^{4}\big)
-{1\over 14400}\big({\rm tr}_{\rm adj}\mathcal{F}^{2}\big)^{3}.
\label{eq:grup_theory_idGS}
\end{align}
The most efficient way to explore the existence of any solution to this condition is by using
group theory identities relating traces of generators in the adjoint representation to those
in the fundamental. We begin with the orthogonal groups~SO($n$), where we have
\begin{align}
{\rm tr}_{\rm adj}\mathcal{F}^{6}&=(n-32){\rm tr}_{\rm f}\mathcal{F}^{6}+15({\rm tr}_{\rm f}\mathcal{F}^{2})
({\rm tr}_{\rm f}\mathcal{F}^{4}), 
\nonumber \\[0.2cm]
{\rm tr}_{\rm adj}\mathcal{F}^{4}&=(n-8){\rm tr}_{\rm f}\mathcal{F}^{4}+3({\rm tr}_{\rm f}\mathcal{F}^{2})^{2},
\label{eq:ids_so(n)}
\\[0.2cm]
{\rm tr}_{\rm adj}\mathcal{F}^{2}&=(n-2){\rm tr}_{\rm f}\mathcal{F}^{2},
\nonumber
\end{align}  
The key lies 
in the first identity. For eq.~\eqref{eq:grup_theory_idGS} to be satisfied it is
necessary that~${\rm tr}_{\rm adj}\mathcal{F}^{6}$ be written as 
a product of traces in the fundamental. This only happens for~SO(32), 
whose adjoint representation also has the right dimension~$N=496$.
In fact, setting~$n=32$ in~\eqref{eq:ids_so(n)}, 
we easily check that~\eqref{eq:grup_theory_idGS} is satisfied. 

Having established that the GS mechanism works for~SO(32), we 
scan for other semisimple groups. The only one for which~${\rm tr}_{\rm adj}\mathcal{F}^{6}$ 
factorizes is~$E_{8}$. The adjoint and fundamental representations of this group coincide, and
we have
\begin{align}
{\rm tr}_{\rm adj}\mathcal{F}^{6}&={1\over 7200}({\rm tr}_{\rm adj}\mathcal{F}^{2})^{3}, 
\nonumber \\[0.2cm]
{\rm tr}_{\rm adj}\mathcal{F}^{4}&={1\over 100}({\rm tr}_{\rm adj}\mathcal{F}^{2})^{2}.
\label{eq:trace_ids_E8}
\end{align} 
Moreover, the adjoint representation of~$E_{8}$ has dimension~$N=248$, 
so~$E_{8}\times E_{8}$ has the correct value of~$N$.
Using~${\rm tr}_{\rm adj}\mathcal{F}_{E_{8}\times E_{8}}^{n}
={\rm tr}_{\rm adj}\mathcal{F}^{n}_{E_{8}}+{\rm tr}_{\rm adj}\mathcal{F}'^{n}_{E_{8}}$,
we confirm that~\eqref{eq:grup_theory_idGS} holds also for~$E_{8}\times E_{8}$. 

Besides~SO(32) and~$E_{8}\times E_{8}$, there is the somewhat trivial solution 
provided by the group~U(1)$^{496}$. In this case fermions, being in the adjoint representation,  
do not couple to the gauge fields, so the only contribution 
to the anomaly comes from the gravitational sector. Removing all terms containing
Chern characters from the anomaly polynomial~\eqref{eq:anomaly_typeI}, 
we verify its factorization implying that anomalies are canceled by the GS mechanism.
A slightly more interesting group is~$\mbox{U(1)}^{248}\times E_{8}$, where 
fermions do couple to the gauge sector through the non-Abelian factor.
Taking into account that~${\rm tr}_{\rm adj}\mathcal{F}^{2n}={\rm tr}_{\rm adj}\mathcal{F}^{2n}_{E_{8}}$,
we use~\eqref{eq:trace_ids_E8} to check that~\eqref{eq:grup_theory_idGS} is satisfied.

With this we exhaust all possibilities. Anomalies in ten-dimensional $\mathcal{N}=1$ SUGRA
coupled to~$\mathcal{N}=1$ SYM can be canceled through the GS mechanism for just {\em four} gauge 
groups: SO(32),~$E_{8}\times E_{8}$,~$\mbox{U(1)}^{496}$, and~$\mbox{U(1)}^{248}\times E_{8}$.
Since classical 
type-I strings only accomodate~SO($2n$), USp($n$), and~U($n$), 
we conclude that~SO(32) is the unique choice of the gauge group making the theory consistent. 
Remarkably, this is also the only group for which one-loop dilaton
tadpoles cancel\footnote{There exists a subtle relation between anomalies and closed string tadpoles: 
spacetime anomalies in type-I string theory are linked to the presence of a nonzero 
tadpole for an unphysical scalar in the R-R sector. 
This tadpole cannot be removed through the Fischler-Susskind mechanism 
and only cancels if the gauge group is SO(32)~\cite{CaiPol}.}~\cite{GS_finitude}.

\paragraph{\bf The physics behind.}

Let us look at the GS mechanism from a more physical angle. 
Besides the restriction on the gauge group just discussed, anomaly cancellation
rests on the existence of a two-form field with the appropriate gauge 
transformation~\eqref{eq:C2_gauge_trans}. The good news for 
type-I strings is that the theory not only allows the ``safe group''~SO(32), but that it also
contains a rank-two antisymmetric tensor~$B_{\mu\nu}$ 
in the Ramond-Ramond (R-R) sector. Its 
kinetic term in the (string frame) low-energy effective action takes the form
\begin{align}
S_{\mbox{\scriptsize type-I}}\supset -{1\over 2(2\pi)^{7}\alpha'^{4}}\int H\wedge\star H,
\label{eq:B-field_kinetic}
\end{align}
where~$4\pi^{2}\alpha'\equiv\ell_{s}^{2}$ is the string length scale.
The GS mechanism can be implemented as outlined above by identifying the properly 
normalized two-form field~$C_{2}$ above with the R-R rank two tensor, that we denote by~$B$.
Its naive field 
strength~$H=dB$ gets modified to\footnote{The gauge transformation of the~$B$-field by~$\omega^{1}_{2}$ 
is in fact also required to preserve local supersymmetry~\cite{BdRdWvN,chapline_manton}.}
\begin{align}
H=dB-{\alpha'\over 4}\left[{\rm tr}_{\rm f}\left(\mathcal{A}d\mathcal{A}+{2\over 3}\mathcal{A}^{3}\right)
-{\rm Tr\,}\left(\omega d\omega+{2\over 3}\omega^{3}\right)\right],
\label{eq:new_field_strength_B}
\end{align} 
where we used the~SO(32) identity~${\rm tr}_{\rm adj}\mathcal{F}^{2}=30\,{\rm tr}_{\rm f}\mathcal{F}^{2}$
[see that last equation in~\eqref{eq:ids_so(n)}] to cancel the denominator 
of~$\lambda=-{1\over 30}$ in~\eqref{eq:HmodifiedGS}.
In addition, we need to add the
GS counterterm~\eqref{eq:GS_counterterm}, which includes the interactions
\begin{align}
S_{\rm ct}&\supset{1\over 768\pi^{5}\alpha'}
\int B\left[{\rm tr}_{\rm f}\mathcal{F}^{4}-{1\over 8}\big({\rm tr}_{\rm f}\mathcal{F}^{2}\big)
\big({\rm Tr\,}\mathcal{R}^{2}\big)
+{1\over 8}{\rm Tr\,}\mathcal{R}^{4}+{1\over 32}\big({\rm Tr\,}\mathcal{R}^{2}\big)^{2}\right].
\label{eq:X8_general10D_traces}
\end{align}
Here we restored the right normalization of the action and wrote again all gauge traces in
the fundamental representation.

\begin{figure}[t]
\centerline{\includegraphics[scale=0.40]{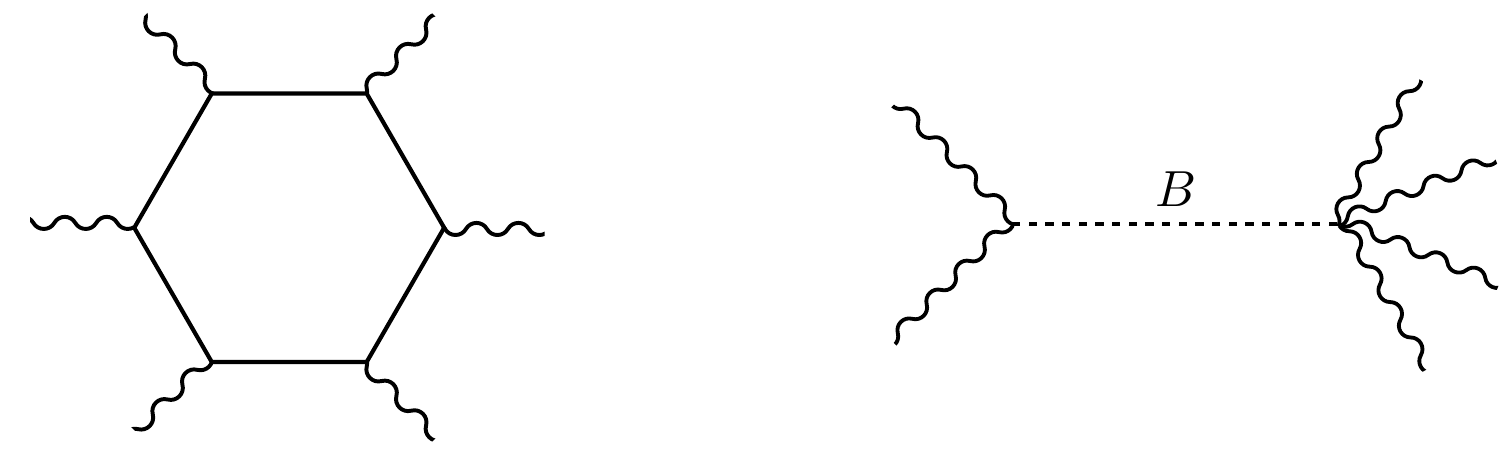}}
\caption[]{Ten-dimensional anomalies are determined by a one-loop hexagon diagram as the one
shown on the left panel of this figure. In the GS mechanism, their contribution to the anomaly is 
canceled by tree-level diagrams as the one depicted on the right panel, 
describing the interchange of a~$B$-field between 
gravitons and/or gauge bosons. Here and in all other diagrams below, the wavy
lines represent either gauge or graviton fields, as explained in the text.}
\label{fig:hexagon_10D}
\end{figure}

These modifications to the low-energy action provide the clue to understanding how
the~$B$-field cancels the anomaly at a diagrammatic level. We know that gauge and gravitational
anomalies in a ten-dimensional field theory come from hexagon diagrams with gravitons and/or gauge bosons
at the vertices (see the left panel of fig.~\ref{fig:hexagon_10D}). As explained in pag.~\pageref{page:hexagon_10D}, there are contributions from all diagrams containing
an even number of gravitons and gauge bosons. Now, the presence of the Chern-Simons forms~$\omega^{0}_{3}$ and~$\Omega^{0}_{3}$
in the field strength~\eqref{eq:new_field_strength_B} induces 
new interactions from the kinetic term~\eqref{eq:B-field_kinetic}: vertices 
with a single~$B$-field and either two gravitons or two gauge bosons.  
Furthermore, the part 
of the GS counterterm shown in~\eqref{eq:X8_general10D_traces} 
introduces three additional five-point vertices. 
They contain one~$B$-field and either four gravitons, four gauge
bosons or two gravitons and two gauge bosons. 

Due to these new interactions there are additional diagrams contributing to the anomaly. In particular
the one shown on the right of fig.~\ref{fig:hexagon_10D}, where either two gravitons
or two gauge bosons combine into a~$B$-field that decays into either four gravitons, four gauge bosons or
two gravitons and two gauge bosons. The GS mechanism works because these tree-level diagrams 
cancel the parity-violating contribution from the hexagons, rendering the theory anomaly-free.
Notice that the source of parity violation in the tree diagram is entirely in 
the right vertex\footnote{Being an antisymmetric tensor, 
the coupling of the $B$-field to four massless bosons contains a Levi-Civita tensor that is however absent from the trivalent vertex. 
As a consequence, the tree-level diagram on the right of fig.~\ref{fig:hexagon_10D} 
violates parity.}. The
coupling of two gravitons or two gauge bosons and a~$B$-field preserves parity due to presence of
the Hodge dual in the kinetic term~\eqref{eq:B-field_kinetic}.

This diagrammatic interpretation brings forward the unusual 
character of the GS mechanism. Anomaly cancellation
in quantum field theory usually proceeds by including either additional species
or new couplings (or both), so 
the offending one-loop diagrams are canceled by other one-loop diagrams with either these extra states 
running in the 
loop or (and) new couplings showing up in the vertices. 
Not so in the GS mechanism. Here we have 
tree-level diagrams cancelling loops\footnote{The closest analogy to the GS cancellation mechanism
has to be found in the Wess-Zumino effective Lagrangian, 
where triangle anomalies are compensated by the tree-level transformation 
of the Nambu-Goldstone bosons~\cite{WZ_lagrangian}.}.

To understand why this happens, we go back to a comment on page~\pageref{page:IRpole} where we pointed out
that anomalies can be seen as arising from 
IR poles in the expectation value of {\em current}, as opposed to UV ambiguities in the 
expectation values of the {\em current divergence}. The diagrams we are discussing 
contribute to the expectation value
of the gauge current and the energy-momentum tensor, so anomalies 
are read from the residue of the parity-violating part of the hexagon diagram in the limit in which the invariant mass of
the~$i$-th and~$j$-th insertions approaches zero, $s_{ij}=-2p_{i}\cdot p_{j}\rightarrow 0$.  
On the other hand, 
since the~$B$-field is massless, its interchange in 
tree diagram on the right of fig.~\ref{fig:hexagon_10D} generates a pole at 
zero-momentum transfer.

For the tree-level diagram to cancel the right poles from the hexagon 
a number of things have to conspire. The~$B$-field is neutral and cannot transfer gauge charge from one
vertex to the other. This means that no cancellation can take place on any pole proportional to a single trace over gauge generators, 
a situation that we avoided by restricting the gauge group to those for which these traces factorize. Similarly, lacking Lorentz indices, it cannot cancel poles 
proportional to~${\rm Tr\,}\mathcal{R}^{3}$ either. These are the terms that we got rid of in the hexagon
by setting~$N=496$.

\paragraph{\bf Type-I superstrings and beyond.}
The modifications to the field theory action of ten-dimensional~$\mathcal{N}=1$ gauged SUGRA 
imposed by anomaly cancellation
do not pose any problems within a field-theoretic context. Indeed, in quantum field theory 
we are always allowed the freedom of ``building'' a Lagrangian
so it fits our low energy requirements. Any couplings we might need to introduce are 
swept under the rug of an eventual UV completion of the theory. 
There is however a problem when the theory we deal with describes the low-energy 
dynamics of some string model. Since 
string theory is UV complete, both the massless states and their couplings are fully determined. Our 
playground is thus pretty much constrained.

This is why we need to go a step further and check whether the GS mechanism is in fact {\em fully implemented} in type-I string theory. 
In other words, it is not enough having the appropriate antisymmetric tensor field in the spectrum, but the new coupling 
stemming from the modified field strength~\eqref{eq:new_field_strength_B} and the GS 
counterterm proportional to~$BX_{8}$ should actually follow from 
the low-energy limit of the string interactions
up to the last factor.

\begin{figure}[t]
\centerline{\includegraphics[scale=0.35]{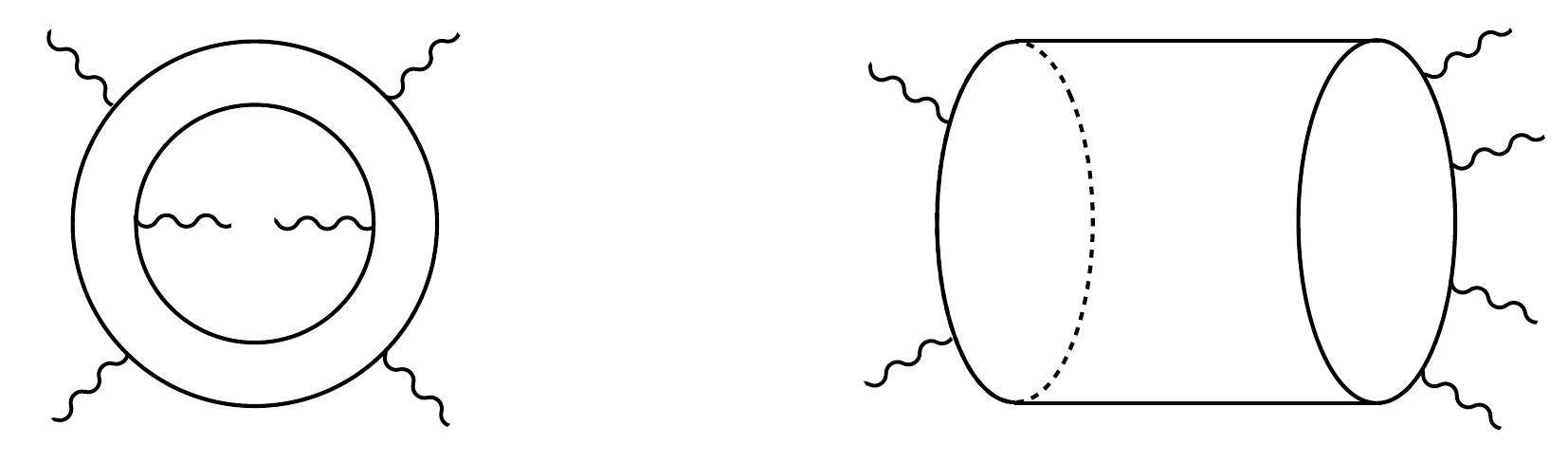}}
\caption[]{In type-I string theory the contribution of 
the nonplanar six-point annulus diagram on the left is recast in the closed-string channel
as the tree-level diagram on the right. Closed string states, among them the R-R massless $B$-field, propagates along the cylinder resulting in the low-energy 
limit in the diagram shown on the right of fig.~\ref{fig:hexagon_10D}.}
\label{fig:string_diags}
\end{figure}

Let us focus on pure gauge anomalies. 
There are three kinds of one-loop string diagrams contributing to them:
the planar orientable annulus with all Yang-Mills vertex operators attached to a single boundary, the nonorientable M\"obius strip 
with six gauge insertions on its boundary, and the nonplanar annulus with four vertices attached to one boundary
and two to the other (remember that gauge bosons belong to the open string sector and therefore couple to the diagram boundaries). 
An explicit calculation~\cite{GS2} shows that for~SO(32) gauge anomalies cancel among the first two topologies,
whose group theory factors are in both cases single traces~${\rm tr}_{\rm adj}(\lambda^{a_{1}}\ldots\lambda^{a_{6}})$.
As to the nonplanar diagram shown on the left 
of fig.~\ref{fig:string_diags}, it is proportional to~${\rm tr}_{\rm adj}(\lambda^{a_{1}}\lambda^{a_{2}})
{\rm tr}_{\rm adj}(\lambda^{a_{3}}\ldots\lambda^{a_{6}})$ and does not contribute to the anomaly. 

This proves that gauge anomalies cancel in full-fledged type-I string theory.
To connect with the field-theoretic analysis of the GS mechanism, we need to look for the IR poles
associated with the three diagrammatic topologies mentioned above. The two first topologies (the planar annulus and
the M\"obius strip) produce the 
poles associated to the field theory hexagon. Surprisingly, the nonplanar annulus
on the left of fig.~\ref{fig:string_diags} also gives a massless pole. Switching from 
the open to the closed string channel, this diagram is transformed into one 
in which the gauge bosons interact with each other through the interchange of a closed string,
as shown on the right of fig.~\ref{fig:string_diags}. 
Close to the zero-momentum transfer, 
the amplitude is dominated by the boundary of the moduli space corresponding 
to a very long cylinder. Moreover, its parity-violating part is nonzero
if the state running along the cylinder is the R-R rank-two antisymmetric tensor. 
For~SO(32), the single trace in the planar annulus and M\"obius strip topologies factorizes and its pole 
is canceled by the cylinder diagram, as required for the GS mechanism to work. 

This proves that the GS mechanism is automatically implemented in type-I string theory, at least for gauge anomalies. 
The analysis of gravitational and mixed anomalies requires more work. Besides the topologies already considered, now
with graviton vertex operators inserted in their interior, we need to compute the
contributions of the torus and the Klein bottle, 
both with and without a boundary. Their explicit evaluation~\cite{open_gen_anom_anal} shows that all anomalies cancel at one loop. The interpretation of the tree-level diagram is now a little bit different. The relevant
parity-violating pole comes again from the~$B$-field propagating along a long cylinder. 
However, depending on the amplitude
there are three separate
possibilities: the cylinder has two gauge boson insertions on one of its boundaries while 
the other one is attached 
to a torus with four graviton vertex operators; it  
joints a sphere with two gravitons to a torus with four gravitons (cf. the diagram on 
the right of fig.~\ref{fig:torus_het} below);
or it connects a sphere with two gravitons to 
a torus with a boundary and two gauge bosons attached to it, while two graviton vertex operators 
are also inserted in its bulk.

It is difficult to exaggerate the importance of the discovery of the 
GS mechanism in the historical development of 
string theory. After the doubts sown by the results of~\cite{AG_Witten} concerning the consistency of
type-I string theory, the discovery of a highly nontrivial cancellation mechanism meant a very 
important push to the theory, so important that with it the first superstring revolution was 
initiated. One of the consequences was the attempts to accommodate the second 
``safe'' group~$E_{8}\times E_{8}$ into the framework of string theory. This led to the formulation of
the heterotic string~\cite{princeton_quartet}, the model that dominated string phenomenology
until the advent of D-branes at the onset of the second superstring revolution. 

The low-energy dynamics of the heterotic string is also that of~$\mathcal{N}=1$ SUGRA coupled
to~$\mathcal{N}=1$ SYM, although in a slightly modified fashion from the one we encountered
for type-I strings~\cite{chapline_manton}. This notwithstanding, 
the theory contains the crucial rank-two antisymmetric tensor field, this time
in the NS-NS sector, and the implementation of the GS mechanism follows the same steps outlined above.  
The absence of an open string sector means that in the heterotic string the calculation of the anomaly only involves 
a single diagram, a torus with six graviton/gauge boson vertex operators as the one shown on the left of~\ref{fig:torus_het}. 
The key to the cancellation of spacetime anomalies in the heterotic string lies in modular 
invariance\footnote{Modular invariance is also at the heart of anomaly cancellation 
in type-II string theory~\cite{type-II_cancellation}.}~\cite{heterotic_anomalies}, 
the very same symmetry restricting the allowed gauge groups
to~SO(32) and~$E_{8}\times E_{8}$. 

Anomalies in the heterotic string are associated with contributions
coming from the boundary of the moduli space of tori with six punctures. If the theory is modular invariant,
this boundary has two components whose contributions cancel each other out: 
the limit~$\tau\rightarrow i\infty$, corresponding to very long
tori, and the limit in which the $i$-th and $j$-th punctures collide. In the first case, the diagram
is dominated by massless states running along the torus and its parity-violating piece gives the 
pole associated with the field theory hexagon diagram. The second component corresponds to the factorization 
limit where the punctured torus degenerates into a sphere containing the two colliding vertex operators
joined by a long cylinder to a torus with the remaining insertions (see the drawing on the right
of fig.~\ref{fig:torus_het}). The associated parity-violating pole 
comes from the propagation of the~$B$-field along the cylinder, 
thus implementing
the GS cancellation mechanism in the low-energy field theory.

\begin{figure}[t]
\centerline{\includegraphics[scale=0.35]{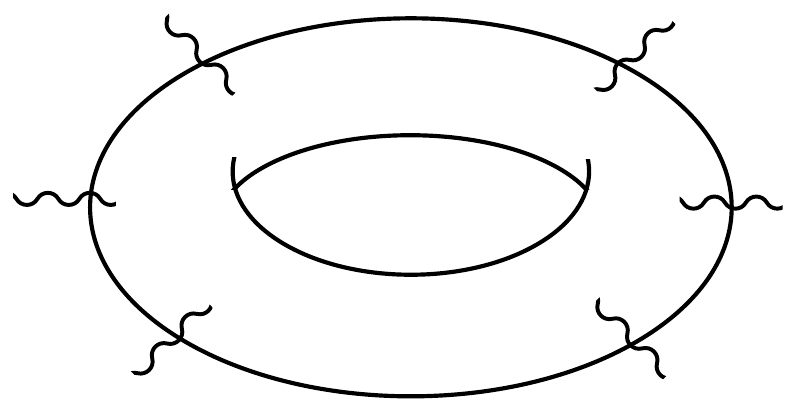}\hspace*{1cm}\includegraphics[scale=0.5]{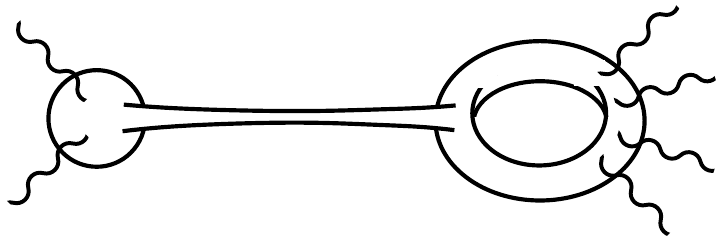}}
\caption[]{All gauge and gravitational anomalies in the heterotic string stem from the torus diagram shown on the left. 
Low-energy poles
from the field theory hexagon
are canceled by the ones coming from  
massless closed string states running in the long propagator joining the sphere and the torus
in the diagram on the right. This diagram also occurs in the calculation of the pure gravitational
anomaly of type-I string theory.}
\label{fig:torus_het}
\end{figure}

\begin{svgraybox}
\centerline{\bf Some further examples of the GS mechanism}\\

The GS mechanism, first identified in the context of type-I string theory, has found implementation in
various secenarios. Here we discuss 
three particular instances of this anomaly cancellation mechanism at work.\\

\noindent
{\bf The ${\bf \mathrm{\bf SO}(16)\boldsymbol{\times}\mathrm{\bf SO}(16)}$ nonsupersymmetric string.}
The $\mathrm{SO}(16)\times\mathrm{SO}(16)$ heterotic string~\cite{SO(16)} is a ten-dimensional 
nonsupersymmetric and tachyon-free fermionic model.
Its spectrum, besides the universal gravity
multiplet containing the graviton, dilaton, and antisymmetric rank-two tensor field, includes
a positive and a negative chirality Majorana-Weyl fermions respectively
transforming in the~$(\boldsymbol{16},\boldsymbol{16})$ 
and $(\boldsymbol{128},\boldsymbol{1})\oplus
(\boldsymbol{1},\boldsymbol{128})$ representations of
$\mathrm{SO}(16)\times\mathrm{SO}(16)$, 
with~$\boldsymbol{16}$ and~$\boldsymbol{128}$ the fundamental
and spinor representations of~SO(16).
Since these are the only fields contributing to the anomaly, 
we can write the relevant anomaly polynomial as
\begin{align}
I_{12}={1\over 2}I_{1\over 2}[\mathcal{F},\mathcal{R}]_{(\boldsymbol{16},\boldsymbol{16})}
-{1\over 2}I_{1\over 2}[\mathcal{F},\mathcal{R}]_{(\boldsymbol{128},\boldsymbol{1})}
-{1\over 2}I_{1\over 2}[\mathcal{F},\mathcal{R}]_{(\boldsymbol{1},\boldsymbol{128})},
\end{align}
where once more the ${1\over 2}$~factors account for the number
of degrees of freedom of a Majorana-Weyl spinor and the subscripts remind us what representation we should
use to evaluate the traces in the corresponding Chern characters. It is a peculiarity of this model that, unlike 
the type-I superstring, we have
fermions transforming in various representations of the gauge group. Interestingly, looking at 
eq.~\eqref{eq:I1/2_12} we see that all pure gravitational anomaly terms cancel among each other, 
due to the identity~$16^{2}-128-128=0$.

We are left then with pure and mixed gauge anomalies.  
In order to check their cancellation it is convenient
to write all Chern characters using traces in the fundamental representation~$\boldsymbol{16}$ of~SO(16).
Remarkably, when this is done~\cite{SO(16)} we find the factorized result
\begin{align}
I_{12}&=(\mbox{ch}_{2}+\mbox{ch}_{2}'-p_{1})X_{8},
\end{align}
where the prime indicates the second factor of~SO(16) and~$X_{8}$ is explicitly given by
\begin{align}
X_{8}={1\over 2}\big(\mbox{ch}_{4}+\mbox{ch}'_{4}\big)-{1\over 48}\big(\mbox{ch}_{2}^{2}+\mbox{ch}_{2}'^{2}-\mbox{ch}_{2}\mbox{ch}'_{2}\big).
\end{align}
Since the $\mbox{SO(16)}\times\mbox{SO(16)}$ heterotic string contains a rank-two antisymmetric tensor in 
its gravity multiplet, we can repeat the analysis presented above for the type-I string 
and conclude that all anomalies cancel
through the GS mechanism.~$\blacksquare$

\noindent
{\bf Type-I string theory with space-filling D- and anti-D-branes.} We consider next
the GS cancellation mechanism for type-IIB strings in the presence of one orientifold O9-plane and 
a stack of~$m$
D9-branes and~$n$ anti-D9-branes~\cite{sugimoto}. The gauge group is~$\mbox{SO}(m)\times\mbox{SO}(n)$
and the charged massless chiral fields are a positive chirality 
fermion in the adjoint of~SO($m$), a positive chirality fermion in 
the symmetric tensor representation of~SO($n$), and
a negative chirality spinor in the bifundamental of~$\mbox{SO}(m)\times\mbox{SO}(n)$. 
These are excitations of open strings with endpoints attached to two D9-branes, 
two anti-D9-branes, and one D9-brane and one anti-D9-brane respectively. 
Adding to this the contributions of the positive helicity gravitino and the negative helicity 
dilatino, the 12-form anomaly polynomial is given by
\begin{align}
I_{12}&=-{1\over 2}I_{1\over 2}[\mathcal{R}]+{1\over 2}I_{3\over 2}[\mathcal{R}]
+{1\over 2}I_{1\over 2}[\mathcal{F},\mathcal{R}]_{({\rm adj},\boldsymbol{1})}
\nonumber \\[0.2cm]
&+{1\over 2}I_{1\over 2}[\mathcal{F},\mathcal{R}]_{(\boldsymbol{1},{\rm sym})}
-{1\over 2}I_{1\over 2}[\mathcal{F},\mathcal{R}]_{({\rm f},{\rm f})},
\end{align}
where the explicit expression for each term can be found in eqs.~\eqref{eq:I1/2_grav}, 
\eqref{eq:I3/2_grav}, and~\eqref{eq:I1/2_12}.
We just look at the two terms spelling trouble for the implementation of the GS mechanism,
which are the ones containing~$p_{3}$ and~$\mbox{ch}_{6}$. The first one is proportional to
\begin{align}
496-{1\over 2}m(m-1)-{1\over 2}n(n+1)-mn=496-{1\over 2}(m-n)(m-n-1),
\end{align}
which cancels for~$m-n=32$. As to the second offending term, we search for identities relating the 
traces of~$\mathcal{F}^{6}$ in the various representations to those in the fundamental. The relevant ones are
\begin{align}
{\rm tr}_{({\rm adj},\mathbf{1})}\mathcal{F}^{6}&=(n-32){\rm tr}_{\rm f}\mathcal{F}^{6}
+15\big({\rm tr}_{\rm f}\mathcal{F}^{2}\big)\big({\rm tr}_{\rm f}\mathcal{F}^{4}\big) ,\nonumber \\[0.2cm]
{\rm tr}_{(\mathbf{1},{\rm sym})}\mathcal{F}^{6}&=
(m+32){\rm tr}_{\rm f}\mathcal{F}'^{6}+15\big({\rm tr}_{\rm f}\mathcal{F}'^{2}\big)\big({\rm tr}_{\rm f}\mathcal{F}'^{4}\big), \\[0.2cm]
{\rm tr}_{({\rm f},{\rm f})}\mathcal{F}^{6}&=n{\rm tr}_{\rm f}\mathcal{F}^{6}+m{\rm tr}_{\rm f}\mathcal{F}'^{6},
\nonumber
\end{align} 
where the primes indicate the field strength associated to the~SO($n$) factor.
Adding these three contributions with their respective signs, we find a cancellation of 
the irreducible terms proportional
to~${\rm tr\,}\mathcal{F}_{1}^{6}$ and~${\rm tr\,}\mathcal{F}_{2}^{6}$. This ensures the 
factorization of the anomaly polynomial
\begin{align}
I_{12}=\left(p_{1}-{1\over 30}\mbox{ch}_{2}-{1\over 30}\mbox{ch}'_{2}\right)Y_{8}
\end{align}
where
\begin{align}
Y_{8}&=-{1\over 48}\left[
{\rm ch}_{4}-{\rm ch}'_{4}
-{1\over 60}p_{1}\big({\rm ch}_{2}-{\rm ch}'_{2}\big)
+{3\over 8}p_{1}^{2}-{1\over 2}p_{2}
\right],
\end{align}
and the prime again indicates the Chern characters associated with~SO($n$).
All gauge, gravitational, and mixed
anomalies thus cancel for~$m-n=32$ through the GS mechanism. As in the case of the type-I string, 
we also need to modify the antisymmetric tensor field strength
\begin{align}
H=dC_{2}-{1\over 30}\omega_{3}^{0}-{1\over 30}\omega'^{0}_{3}+\Omega_{3}^{0},
\end{align}
and add the corresponding GS counterterm to the action
\begin{align}
\Gamma_{\rm ct}=c\int_{\mathcal{M}_{10}}C_{2} Y_{8}
+{1+\alpha\over 2}c\int_{\mathcal{M}_{10}}\left({1\over 30}\omega_{3}^{0}+{1\over 30}\omega'^{0}_{3}
-\Omega^{0}_{3}\right)Y_{7}^{0},
\end{align}
with~$Y_{8}=dY_{7}^{0}$.$\blacksquare$

\noindent
{\bf A four-dimensional version of the GS mechanism.}
As a final example, we study the implementation of the GS mechanism in a purely field-theoretical
setup. Let us consider the
theory of a positive chirality Weyl fermion coupled to a U(1) gauge field propagating, for simplicity,
on flat spacetime. The corresponding anomaly polynomial is
\begin{align}
I_{6}=\mbox{ch}_{3},
\label{eq:anomaly_pol_I6_axion}
\end{align}
and the theory has an anomaly given by
\begin{align}
\delta_{\eta} S_{\rm eff}&=2\pi i\int \omega^{1}_{4}
=-\int \eta\mbox{ch}_{2},
\label{eq:axion_anomaly}
\end{align}
where we have used that~$\omega^{1}_{4}$ is proportional to~$\mathcal{F}^{2}$
and therefore to the second
Chern character.

For a~U(1) gauge group, the anomaly polynomial~\eqref{eq:anomaly_pol_I6_axion}
trivially factors
\begin{align}
\mbox{ch}_{3}={1\over 3}\mbox{ch}_{1}\mbox{ch}_{2},
\end{align} 
and the GS mechanism is implemented by an 
axion field~$\theta$ with the gauge transformation
\begin{align}
\delta_{\eta}\theta=m_{\rm GS}\eta.
\end{align}
and a gauge-invariant kinetic term 
\begin{align}
S \supset {1\over 2}\int \big(d\theta-m_{\rm GS}\mathcal{A}\big)\wedge \star 
\big(d\theta-m_{\rm GS}\mathcal{A}\big).
\label{eq:kinetic_term_axion}
\end{align}
Here,~$m_{\rm GS}$ is a constant with dimensions of mass. 
The anomaly is then canceled by the addition of
the four-dimensional GS counterterm
\begin{align}
S_{\rm ct}=-{1\over m_{\rm GS}} \int \theta \mbox{ch}_{2}
={1\over 8\pi^{2}m_{\rm GS}}\int \theta\mathcal{F}^{2}.
\label{eq:axion_ct}
\end{align}
Indeed,~$\delta_{\eta}S_{\rm eff}+\delta_{\eta}S_{\rm ct}=0$ and gauge invariance is 
preserved quantum-mechanically.

This cancellation can also be understood in diagrammatic terms by noticing 
that~\eqref{eq:kinetic_term_axion} introduces a kinetic mixing between the axion and the gauge field
of the form~$\mathcal{A}^{\mu}\partial_{\mu}\theta$.
On the other hand, the counterterm~\eqref{eq:axion_ct} induces a trivalent vertex with one axion
and two gauge fields. Thus, the triangle diagram producing the anomaly~\eqref{eq:axion_anomaly} is canceled by 
a tree-level diagram where a gauge field transmutes into an axion that then decays into two other
gauge fields. This is the four-dimensional analog of the diagram shown on the right panel
of fig.~\ref{fig:hexagon_10D}. In fact, the axion is just a two-form field in disguise, since both
fields are related by Hodge duality in four 
dimensions,~$d\theta=\star dB$~\cite{bonnefoy_dudas}.~$\blacksquare$

\end{svgraybox}

\section{Anomaly inflow}

Having focused so far on anomaly cancellation, our discussion has avoided
the evaluation of currents. In the gauge case, the (consistent) current is obtained 
by taking variations of the nonlocal quantum effective
action~$\Gamma_{\rm eff}$ with respect to the gauge one-form~$\mathcal{A}$. 
The result splits into two pieces, 
one given by an integral over the ``bulk''~$\mathcal{M}_{2n-1}$ space
and a second defined on its~$(2n-2)$-dimensional boundary\footnote{A direct way of getting
this result is by applying the generalized transgression formula of~\cite{MSZ} to the 
gauge Chern-Simons form~$\omega^{0}_{2n-1}$, using the one-parameter family of connections
defined by~$\mathcal{A}_{t}=\mathcal{A}+t\delta A$ (see~\cite{MMVVM18} for details).}
\begin{align}
\delta\Gamma_{\rm eff}=\int_{\mathcal{M}_{2n-1}}{\rm Tr\,}\big(J_{\rm bulk}\delta\mathcal{A}\big)
+\int_{\partial\mathcal{M}_{2n-1}}{\rm Tr\,}\big(X\delta\mathcal{A}\big).
\label{eq:deltaGamma_effdeltaA}
\end{align}
The quantities inside the traces in both integrals are explicitly given by
\begin{align}
J_{\rm bulk}&={i^{n}\over (n-1)!(2\pi)^{n-1}}\mathcal{F}^{n-1}, \nonumber \\[0.2cm]
X&={i^{n}\over (n-1)!(2\pi)^{n-1}}\int_{0}^{1}tdt\,\Big(\mathcal{F}_{t}^{n-2}\mathcal{A}
+\mathcal{F}_{t}^{n-3}\mathcal{A}\mathcal{F}_{t}+\ldots+\mathcal{A}\mathcal{F}_{t}^{n-2}\Big),
\label{eq:X_def}
\end{align}
with~$\mathcal{F}_{t}\equiv t\mathcal{F}+t(t-1)\mathcal{A}^{2}$. We particularize~\eqref{eq:deltaGamma_effdeltaA}
to the case of infinitesimal gauge 
transformations, $\delta_{\chi}\mathcal{A}=d\chi+[\mathcal{A},\chi]\equiv D\chi$,
where~$D$ denotes the gauge covariant derivative. 
Applying the Stokes theorem, we find the 
gauge variation of the effective action
\begin{align}
\delta_{\chi}\Gamma_{\rm eff}=-\int_{\mathcal{M}_{2n-1}}{\rm Tr\,}\big(\chi D J_{\rm bulk}\big)
+\int_{\partial\mathcal{M}_{2n-1}}{\rm Tr\,}\Big[\chi\big( J_{\rm bulk}-DX\big)\Big].
\label{eq:deltachiGamma_effbulk+bdy1}
\end{align}
As we know, the left-hand side of this equation gives the consistent anomaly [cf.~\eqref{eq:gauge_anomaly_def1}], 
which we express in terms of the consistent current~$J_{\rm cons}$ as\footnote{To simplify the discussion, 
we work for the time being 
with the Hodge duals of all one-form currents.}
\begin{align}
\delta_{\chi}\Gamma_{\rm eff}\equiv-\int_{\partial\mathcal{M}_{2n-1}}{\rm Tr\,}\big(\chi DJ_{\rm cons}\big).
\end{align}
As to the right-hand side of~\eqref{eq:deltachiGamma_effbulk+bdy1}, 
the first thing to notice is that the Bianchi identity~$D\mathcal{F}=0$
implies~$DJ_{\rm bulk}=0$ so the first, nonlocal term vanishes. 
In addition to this, the quantity~$X$ defined in~\eqref{eq:X_def} is
minus the Bardeen-Zumino term~\cite{BZ_NPB} relating the consistent and covariant 
currents,~$J_{\rm cov}=J_{\rm cons}-X$
(see our discussion in page~\pageref{pag:BZ_term}). We thus obtain a very suggestive
expression
\begin{align}
DJ_{\rm cov}=- J_{\rm bulk}\bigg|_{\partial\mathcal{M}_{2n-1}},
\label{eq:anomaly_inflow1}
\end{align}
stating that the covariant anomaly equals minus the bulk current evaluated at the 
boundary.

To understand the implications of this relation, let us take a step back and 
review what we have done. By expressing it as an integral over a higher-dimensional space, 
the {\em nonlocal} effective action~$\Gamma_{\rm eff}$ can be
interpreted as describing the {\em local} quantum effective dynamics of gauge fields interacting with 
Dirac fermions in the odd-dimensional bulk Euclidean
spacetime~$\mathcal{M}_{2n-1}$. This theory is free of gauge anomalies, 
as it is manifest in the fact that the quantum 
current~$J_{\rm bulk}$ is conserved,~$DJ_{\rm bulk}=0$.
The situation is quite different on the even-dimensional boundary theory, which 
contains chiral fermions and where~$DJ_{\rm cov}$ (or~$DJ_{\rm cons}$ for that matter)
has a nonvanishing value, 
signaling the existence of a gauge anomaly. 
Equation~\eqref{eq:anomaly_inflow1} provides the clue to understand physically what is going on:
the gauge anomaly results from the {\em inflow} of gauge charge from the bulk onto the boundary,
as shown by the fact that the value of the bulk current there precisely cancels the rate of 
nonconservation of the gauge charge. 

This general argument 
illustrates the basic features of the mechanism of anomaly inflow first pointed out 
in~\cite{CH} (see also~\cite{harvey_review} for a review).
Anomalous theories can be embedded in higher-dimensional, anomaly-free theories
so that violations in the conservation of charge are accounted for by the inflow of 
charge from the higher-dimensional bulk. Or the other way around, nonanomalous theories
in the presence of topological defects may have anomalies supported on their worldvolumes, 
which are nevertheless canceled by the charge flowing from outside the defect.

\begin{svgraybox}
\centerline{\bf A simple example of anomaly inflow}\\
\label{pag:axion_string}

\nopagebreak

Let us see the workings of anomaly inflow in a particular example, that of 
a massless Dirac fermion in 3+1 dimensions in the presence of an axion string defect~\cite{CH,BH,harvey_review}. 
The relevant terms in the action are
\begin{align}
S=\int d^{4}x\,\Big[i\overline{\psi}\gamma^{\mu}\partial_{\mu}\psi+\overline{\psi}\big(\varphi_{1}
+i\gamma_{5}\varphi_{2}\big)\psi\Big].
\end{align}
Here, $\varphi\equiv \varphi_{1}+i\varphi_{2}$ is a complex scalar field 
in the vacuum configuration
\begin{align}
\varphi(x)&=f(\rho)e^{i\theta(x)},
\label{eq:soliton_conf}
\end{align}
where~$\rho^{2}=(x^{1})^{2}+(x^{2})^{2}$ and~$f(\rho)$ satisfies~$f(\rho\rightarrow 0)=0$
and~$f(\rho\rightarrow\infty)=v$. This describes a stringlike vortex localized along the~$x^{3}\equiv z$
direction.

The massless Dirac fermion contains two opposite chiralities and can be coupled to 
an electromagnetic U(1) without gauge anomalies. The question however is whether 
the string defect supports chiral fermion modes that might induce anomalies on its
worldvolume. To clarify the issue, we split the Dirac fermion into its two 
chiralities~$\gamma_{5}\psi_{\pm}=\pm\psi_{\pm}$ and separate
the string worldvolume 
directions~$x^{a}\equiv(x^{0},x^{3})$
from the transverse coordinates~$(x^{1},x^{2})=(\rho\cos\phi,\rho\sin\phi)$. 
We also factorize the four-dimensional chirality matrix~$\gamma_{5}=i\gamma^{0}\gamma^{1}\gamma^{2}\gamma^{3}$
as~$\gamma_{5}=\overline{\gamma}\gamma_{T}$, where $\overline{\gamma}=-\gamma^{0}\gamma^{3}$ is the two-dimensional 
chirality matrix and~$\gamma_{T}\equiv -i\gamma^{1}\gamma^{2}$. 
Doing all this, the Dirac equation is recast as the pair of equations
\begin{align}
i\gamma^{a}\partial_{a}\psi_{\pm}+i\gamma^{2}\big(\cos\phi+i\overline{\gamma}_{T}\sin\phi\big)\partial_{\rho}\psi_{\pm}
=f(\rho)e^{\mp i\phi}\psi_{\mp}.
\end{align}
admitting the solution
\begin{align}
\psi_{-}&=\eta(x^{a})\exp\left[-\int_{0}^{\rho}d\rho'f(\rho')\right], \nonumber \\[0.2cm]
\psi_{+}&=-i\gamma^{2}\psi_{-}.
\end{align}
Here~$\eta$ is a negative chirality spinor satisfying the two-dimensional
Dirac equation on the string worldvolume
\begin{align}
i\gamma^{a}\partial_{a}\eta=0, \hspace*{1cm} \overline{\gamma}\eta=-\eta.
\end{align}

We have shown the existence of a zero mode in the bulk Dirac equation corresponding to a
massless spinor along the string defect whose chirality 
is correlated with its direction of propagation,~$(\partial_{0}-\partial_{3})\eta=0$. As a consequence, when coupling the theory 
to an external electromagnetic field~$\mathcal{A}_{\mu}$, the chiral zero mode triggers a gauge anomaly on the string worldvolume 
\begin{align}
\partial_{a}J^{a}={e\over 4\pi}\epsilon^{ab}\mathcal{F}_{ab}={e\mathcal{E}\over 2\pi},
\label{eq:anomaly_string_WV}
\end{align}
with~$\mathcal{E}$ the external electric field along~$x^{3}$. This expression gives
the amount of charge nonconservation per unit time and unit length.  
Notice that~\eqref{eq:anomaly_string_WV} gives the covariant anomaly, which in two dimensions is
one-half the value of the consistent anomaly (for an explicit calculation of the consistent anomaly in this case, 
see, for example,~\cite{QFT_book}).
Let us recall that this happens on a string defect otherwise embedded in a theory that as a whole
is free from gauge anomalies.

To give a physical picture of what is going on, we study the bulk theory outside the string defect
and compute the one-loop corrected gauge current in the presence of the soliton~$\varphi$. In the adiabatic approximation where
scalar field gradients are small, this is given by the Goldstone-Wilczek current~\cite{GW}
\begin{align}
\langle J^{\mu}\rangle_{\rm bulk}=-{ie\over 16\pi^{2}}\epsilon^{\mu\nu\alpha\beta}{\varphi^{*}\partial_{\nu}\varphi-\varphi\partial_{\nu}\varphi^{*}
\over |\varphi|^{2}}\mathcal{F}_{\alpha\beta}.
\end{align}
Evaluating it on the vacuum solution~\eqref{eq:soliton_conf} 
and in the region where~$f(\rho)$ is well approximated by its asymptotic value, we 
get
\begin{align}
\langle J^{\mu}\rangle_{\rm bulk}={e\over 8\pi^{2}}\epsilon^{\mu\nu\alpha\beta}\partial_{\nu}\theta \mathcal{F}_{\alpha\beta}.
\label{eq:anomaly_string_worldvolume}
\end{align}

To clarify the workings of anomaly inflow in this case, let us take a gauge configuration describing
an electric field~$\mathcal{E}$ along~$x^{3}=z$. 
Thus,~$\mathcal{F}_{ab}=\mathcal{E}\epsilon_{ab}$ with
all remaining components of the field strength equal to zero. The bulk electric current can then be written
as
\begin{align}
\langle\boldsymbol{J}\rangle_{\rm bulk}=-{e\mathcal{E}\over 4\pi^{2}}
(\pmb{\bm{\nabla}}\theta)\times \mathbf{u}_{z}.
\end{align}
Moreover, for an axion string solution we have~$\theta(x)=\phi$, so the previous result takes the simpler form 
\begin{align}
\langle\boldsymbol{J}\rangle_{\rm bulk}=-{e\mathcal{E}\over 4\pi^{2}\rho}\mathbf{u}_{\rho}.
\label{eq:bulk_current_vector}
\end{align}
This shows the existence of radial charge transport from the bulk towards the defect. Computing
the flux of the bulk current~\eqref{eq:bulk_current_vector}
through a cylindrical surface centered on the string, we find the incoming charge 
per unit length and unit time to be
\begin{align}
{dQ\over dtdL}={e\mathcal{E}\over 2\pi},
\end{align}
which exactly reproduces the rate of violation of electric charge conservation on the string worldvolume, 
as given in eq.~\eqref{eq:anomaly_string_worldvolume}. The gauge anomaly on the defect is
thus canceled by the inflow of charge from the bulk.

Our presentation here has closely followed ref.~\cite{CH}. It is possible nevertheless to recast the analysis in 
a language similar to the one used at the beginning of this section. We start with the bulk effective action describing an
axion field coupled to the electromagnetic field
\begin{align}
S_{\rm axion}=-{e^{2}\over 8\pi^{2}}\int_{V}\theta \mathcal{F}^{2},
\end{align}
and take variations with respect to the gauge field
\begin{align}
\delta S_{\rm axion}={e^{2}\over 4\pi^{2}}\int_{V}d\theta\mathcal{F}\delta \mathcal{A}
-{e^{2}\over 4\pi^{2}}\int_{\partial V}\theta\mathcal{F}\delta\mathcal{A}.
\end{align}
Here we assumed that all fields go to zero at infinity and take~$\partial V$ a cylindrical surface surrounding the 
axion string. The expression obtained exhibits the structure displayed in eq.~\eqref{eq:deltaGamma_effdeltaA}, so we identify
\begin{align}
J_{\rm bulk}&={e\over 4\pi^{2}}d\theta\mathcal{F},
\end{align}
where we absorbed a power of~$e$ into~$\delta\mathcal{A}$.
This is equivalent to eq.~\eqref{eq:anomaly_string_worldvolume} and 
using~\eqref{eq:anomaly_inflow1} we retrieve
the two-dimensional covariant anomaly~\eqref{eq:anomaly_string_WV}~\cite{naculich}.~$\blacksquare$

\end{svgraybox}

\paragraph{\bf Anomalous D-brane couplings.}
Due to its universal nature, anomaly inflow is at work in a wide range of physical situations, 
ranging from condensed matter and fluid dynamics to lattice field theory and string theory.
As to the last field, Polchinski's discovery of 
D-branes~\cite{dbranes_polchinski} as the sources of R-R charge
not only solved a long-standing riddle. It also put the focus on the plethora of extended solitonic 
defects present in string theory, many of which had been already studied in the realm of supergravity (see~\cite{DKL} for a contemporary review on the issue). 
It is only natural that anomaly inflow found immediate application in this context. 

In a consistent string theory, any anomaly supported on the worldvolume of 
a defect (D-branes, orientifolds,~NS-branes,...) has to be canceled 
by charge inflow from the bulk. The condition for this to 
happen determines the gauge and gravitational couplings of the extended object~\cite{GHM,CY} (see also~\cite{johnson,szabo} for reviews). 
Let us see how this works in a case that looks very much like a generalization of the axion string discussed in page~\pageref{pag:axion_string}. 
We consider a $(p+1)$-form field~$C_{p+1}$ coupled to a $p$-brane embedded in a~$D$-dimensional flat space~$\mathcal{M}_{D}$.
The relevant terms in the bulk low-energy action are
\begin{align}
S\supset -{1\over 4}\int_{\mathcal{M}_{D}} H_{p+2}\wedge \star H_{p+2}+{\mu_{p}\over 2}
\int_{\mathcal{B}_{p}}C_{p+1},
\end{align}
where~$H_{p+2}=dC_{p+1}$ 
is the gauge-invariant field strength and~$\mathcal{B}_{p}$ the~$(p+1)$-dimensional brane worldvolume.
The associated equations of motion show that~$C_{p+1}$ is sourced by the $p$-brane
\begin{align}
d\star H_{p+2}=\mu_{p}\delta_{D-p-1}(\mathcal{B}_{p+1}\hookrightarrow\mathcal{M}_{D}),
\label{eq:eom_brane_p+1form}
\end{align}
where we introduced a~$(D-p-1)$-form delta function 
satisfying\footnote{These delta functions are known as de Rham currents.}
\begin{align}
\int_{\mathcal{M}_{D}}\delta_{D-p-1}(\mathcal{B}_{p}\hookrightarrow\mathcal{M}_{D})\alpha_{p+1}=\int_{\mathcal{B}_{p}}\alpha_{p+1},
\label{eq:form_delta_funct}
\end{align}
for any~$(p+1)$-form~$\alpha_{p+1}$ defined on~$\mathcal{B}_{p+1}$. A look at eq.~\eqref{eq:eom_brane_p+1form}
indicates that the Hodge dual of this object can be interpreted as a 
worldvolume-supported $(p+1)$-form current~$J_{p+1}$
\begin{align}
\star J_{p+1}=\delta_{D-p-1}(\mathcal{B}_{p}\hookrightarrow \mathcal{M}_{D}),
\end{align}
so the equations of motion for~$C_{p+1}$ are recast as~$d\star H_{p+2}=\mu_{p}\star J_{p+1}$.

Let us take~$p$ to be odd
and assume that the $p$-brane worldvolume supports a chiral fermion zero mode leading 
to gauge anomalies. From eq.~\eqref{eq:integrated_anomaly} we know that the anomalous variation of the brane action is given by
\begin{align}
\delta S_{\rm brane}=2\pi i\int_{\mathcal{B}_{p}}I_{p+1}^{1},
\label{eq:anomalus_var_brane}
\end{align}
where~$dI_{p+1}^{1}=\delta I_{p+2}^{0}$, with~$I_{p+2}^{0}$ the Chern-Simons descended 
from the brane anomaly polynomial~$I_{p+3}$.
If the bulk theory on~$\mathcal{M}_{D}$ is anomaly-free, its action should contain the coupling
\begin{align}
\Delta S_{\rm bulk}={2\pi i\over \mu_{p}}\int_{\mathcal{M}_{D}}\star H_{p+2}\wedge I^{0}_{p+2},
\label{eq:bulk_term_H}
\end{align}
whose gauge variation cancels the anomaly on the brane~\eqref{eq:anomalus_var_brane}
\begin{align}
\delta \Delta S_{\rm bulk}=-{2\pi i\over \mu_{p}}\int_{\mathcal{M}_{D}}d\star H_{p+2}\wedge I_{p+1}^{1}=-
2\pi i\int_{\mathcal{B}_{p+1}}I_{p}^{1}.
\end{align}
In the second identity we have used the equations of motion~\eqref{eq:eom_brane_p+1form}. 
Just by requiring that anomaly inflow restores the consistency of theory we determined the 
coupling in eq.~\eqref{eq:bulk_term_H}. The other way around, a 
computation of the term~\eqref{eq:bulk_term_H}
from the bulk theory determines~$\mu_{p}$, measuring the coupling of the~$p$-brane to~$C_{p+1}$.

We want to apply the philosophy of this calculation to the case of D$p$-branes. We might be tempted to consider the vanilla case of a single flat 
D$p$-brane in flat space (or a parallel stack of them). In type-IIB theory~$p$ takes odd values so the $(p+1)$-dimensional 
D-brane worldvolume is even-dimensional and may contain chiral fermions.
Yet, there is
the same number of positive and negative chirality fermions on the brane, all in the 
adjoint representation of the gauge group. This means that the theory is anomaly-free. 
Another way to understand the absence of gauge brane anomalies in this setup 
is by taking into account that the bulk type-IIB 
theory does not contain gauge fields. 
Were the theory on the brane anomalous, its anomaly would have no chance of
being canceled by accretion/depletion of gauge charge from the ten-dimensional bulk. 

We need to consider less simple configurations including also the effects of gravitational anomalies. One interesting scenario is that of intersecting branes whose combined worldvolumes span the 
whole ten-dimensional spacetime while producing a chiral theory at an even-dimensional intersection~\cite{GHM}. 
Anomalies there are compensated by the inflow of charge from the ``parent'' D-branes worldvolumes. 
The condition for this cancellation to happen determines the anomalous gauge and gravitational 
couplings of the D$p$-brane. 

Another way to generate a chiral theory on a D$p$-brane, with~$p$ odd, is by playing with curvature~\cite{CY}. 
Let us consider 
a stack of $N$ coincident D$p$-branes whose worldvolume wrap~$\mathcal{B}_{p+1}$. In this D-brane background 
the ten-dimensional local Lorentz group~$\mbox{SO}(1,9)$ breaks down to~$\mbox{SO}(1,p)\times\mbox{SO}(9-p)$,
where the first factor is the local Lorentz group on the brane worldvolume and the 
second the $R$-symmetry of the maximal SYM theory describing the dynamics on the brane. 
This breaking of the structure group of the tangent bundle~$T\!\mathcal{M}$ reflects 
its decomposition into a Whitney sum of the brane tangent and normal bundles
\begin{align}
T\!\mathcal{M}=T\!\mathcal{B}_{p}\oplus TN.
\label{eq:whitney_sum}
\end{align}
At the same time, fermions in the~$\boldsymbol{16}$ Majorana-Weyl
representation of~$\mbox{Spin}(1,9)$ split 
into representations of $\mbox{Spin}(1,p)\times \mbox{Spin}(9-p)$
with correlated chiralities.
As an example, for~$p=3$ we have the~$(\boldsymbol{2},\boldsymbol{4})\oplus 
(\overline{\boldsymbol{2}},\overline{\boldsymbol{4}})$ representation of~$\mbox{Spin}(1,3)\times
\mbox{Spin}(6)$, whereas for~$p=5$ the~$\boldsymbol{16}$ of~$\mbox{Spin}(1,9)$
decomposes as~$(\boldsymbol{4}_{\boldsymbol{p}},\boldsymbol{2}_{\boldsymbol{p}})\oplus
(\boldsymbol{4}'_{\boldsymbol{p}},\boldsymbol{2}'_{\boldsymbol{p}})$. Incidentally, 
the previous examples display an interesting general
property: for~$p=1$ mod 4 the two spinor chiralities transform in independent representations
of~$\mbox{Spin}(9-p)$, whereas when~$p=3$ mod 4 they are complex conjugate 
of each other (see, for example,~\cite{seiberg_rev}). 

The group~$\mbox{Spin}(9-p)$ is a gauge invariance of the theory whose gauge
potential is given by the normal space components of the spin connection,~$\omega_{N}$.  
In addition, chiral fermions on the brane also have~U($N$) Chan-Paton quantum numbers. 
Being massless excitations of open string with endpoints lying on a brane of the stack, 
they transform in its~$\mbox{\bf N}\otimes\overline{\mbox{\bf N}}$ representation, 
which is real and therefore ``safe'' from the point of view of anomalies (in other words, 
U($N$) transformations are chirality-blind).
In the mathematical lingo what 
we have is a vector bundle~$(V\otimes S_{N}^{+})\oplus (V\otimes S_{N}^{-})$,
where~$V$ is the Chan-Paton bundle and~$S^{\pm}_{N}$ are the spin bundles over the normal space associated
with each chirality. 

The only source of anomalies on the brane worldvolume are chiral fermions, 
so the anomaly polynomial is given by\footnote{To be clear, there are some subtleties 
if~$p=3$ mod 4 when fermions with opposite chiralities transform in complex conjugate representations
of~$\mbox{Spin}(9-p)$. This notwithstanding, the anomaly polynomial is the same as in the case~$p=1$ mod 4~\cite{CY}.}~[cf.~\eqref{eq:index_spin1/2}]
\begin{align}
I&={1\over 2}\widehat{A}(\mathcal{B}_{p})\big[\mbox{ch}(V\otimes S^{+}_{N})-\mbox{ch}(V\otimes S^{-}_{N})\big] \nonumber \\[0.2cm]
&={1\over 2}\widehat{A}(\mathcal{B}_{p})\mbox{ch}(V)\big[\mbox{ch}(S^{+}_{N})-\mbox{ch}(S^{-}_{N})\big],
\label{eq:I_anomcoup_1}
\end{align} 
where the factor of~${1\over 2}$ in front
accounts for Majorana-Weyl fermions 
and we have implemented the factorization property of the Chern 
character for product vector bundles,~$\mbox{ch}(U\otimes W)=\mbox{ch}(U)\mbox{ch}(W)$.
The Chern character of the Chan-Paton bundle can be further rewritten as
\begin{align}
\mbox{ch}(V)&\equiv \mbox{ch}_{\mathbf{N}\otimes\overline{\mathbf{N}}}(\mathcal{F})
=\mbox{ch}_{\mathbf{N}}(\mathcal{F})\mbox{ch}_{\overline{\mathbf{N}}}(\mathcal{F})
=\mbox{ch}(\mathcal{F})\mbox{ch}(-\mathcal{F}).
\end{align}
In the last equality we used the Hermitian character of the~U($N$) generators and
dropped the subscript with the understanding that all traces are 
computed in the fundamental representation of~U($N$). With this, eq.~\eqref{eq:I_anomcoup_1} is recast as
\begin{align}
I&={1\over 2}\widehat{A}(T\!\mathcal{B}_{p})\mbox{ch}(\mathcal{F})\mbox{ch}(-\mathcal{F})
\big[\mbox{ch}(S^{+}_{N})-\mbox{ch}(S^{-}_{N})\big].
\label{eq:I_anomcoup_2}
\end{align}
Here it is manifest that the anomaly results from the asymmetry between~$S^{+}_{N}$ and~$S^{-}_{N}$. 
It remains to see how this is related to the 
curvature of the normal bundle,~$\mathcal{R}_{N}=d\omega_{N}+\omega_{N}^{2}$.

To achieve this, we need an object that somehow we managed to do without so far, the Euler class. For a manifold~$\mathcal{M}$ of
dimension~$2k$, the Euler class~$e(\mathcal{M})$ is defined as the~$2k$-form
\begin{align}
e(\mathcal{M})=x_{1}\ldots x_{k},
\end{align}
where~$x_{i}$ are again the skew eigenvalues introduced in eq.~\eqref{eq:curvature_matrix}. 
The Euler class is thus the Pfaffian of~${1\over 2\pi}\mathcal{R}^{a}_{\,\,\,b}$ or, 
equivalently, the square root of the 
maximal Pontrjagin class.
With this new polynomial at hand, we apply the general relation (see, for example,~\cite{nash})
\begin{align}
\mbox{ch}(S^{+}_{\mathcal{M}})-\mbox{ch}(S^{-}_{\mathcal{M}})=\prod_{i=1}^{k}\Big(e^{x_{i}/2}-e^{-x_{i}/2}\Big)
={e(\mathcal{M})\over \widehat{A}(\mathcal{M})},
\end{align} 
and rewrite the anomaly polynomial on the brane~\eqref{eq:I_anomcoup_2} as
\begin{align}
I&={1\over 2}\mbox{ch}(\mathcal{F})\mbox{ch}(-\mathcal{F})
{\widehat{A}(T\!\mathcal{B}_{p})\over \widehat{A}(TN)}e(TN).
\label{eq:anomaly_pol_brane}
\end{align}
Using the explicit expression of the $A$-roof in~\eqref{eq:A-roof},
the quotient in the previous equation can be expanded in terms of the Pontrjagin 
classes for the tangent and normal bundle,~$p_{i}(\mathcal{R}_{T})\equiv p_{i}$ and~$p_{i}(\mathcal{R}_{N})\equiv p_{i}'$
\begin{align}
{\widehat{A}(T\!\mathcal{B}_{p})\over \widehat{A}(TN)}&=1-{1\over 24}\big(p_{1}-p_{1}') \nonumber \\[0.2cm]
&+{1\over 5760}\big(7p_{1}^{2}+3p'^{2}_{1}-10p_{1}p_{1}'-4p_{2}+4p_{2}'\big)
+\ldots
\end{align}
Notice how the chiral asymmetry is controlled by the curvature of the normal bundle: 
setting~$\mathcal{R}_{N}=0$ implies~$e(TN)=0$ and the anomaly polynomial vanishes. 

We have completed half our task. 
Brane anomalies are then computed from the polynomial~\eqref{eq:anomaly_pol_brane} following the standard descend method: we extract its $p+3$-form piece~$I_{p+3}=dI_{p+2}^{0}$ and get the anomaly as~$\delta I^{0}_{p+2}
=dI^{1}_{p+1}$. In a consistent theory this has to be canceled by charge inflow from the bulk. 

This is the condition we are imposing
now to determine the couplings of the D$p$-brane to the various R-R fields of type-IIB string theory.
Let us define the field
\begin{align}
C\equiv \sum_{k=0}^{5}C_{2k},
\label{eq:CdefsumCps}
\end{align}
and assume that its dynamics is described by the action
\begin{align}
S_{\rm RR}=-{1\over 4}\int_{\mathcal{M}_{10}} H\wedge \star H-{1\over 2}\sum_{k}\mu_{k}\int_{\mathcal{B}_{k}}CY_{k},
\label{eq:RRaction_ansatz}
\end{align}
where~$H\equiv dC$ is the field strength. In the previous ansatz we have 
introduced coupling constants
$\mu_{k}$ to the different D-branes, 
while~$Y_{k}$ are invariant polynomials built from the brane gauge field strength and the curvature.
Their overall normalization is chosen 
so they can be written as~$Y_{k}=N_{k}+dY_{k}^{0}$, with~$N_{k}$ the number of D-branes wrapping
on the worldvolume~$\mathcal{B}_{k}$. 
After an integration by parts, 
we have
\begin{align}
S_{\rm RR}= -{1\over 4}\int_{\mathcal{M}_{10}} H\wedge \star H
-{1\over 2}\sum_{k}\mu_{k}\int_{\mathcal{M}_{10}}\delta_{\,9-k}(\mathcal{B}_{k}\hookrightarrow \mathcal{M}_{10})\big(N_{k}C-HY_{k}^{0}\big),
\label{eq:RRaction}
\end{align}
where to express the second
term on the right-hand side as an integral over the ten-dimensional bulk spacetime,
we have used the~$(9-k)$-form delta function introduced in~\eqref{eq:form_delta_funct}.

The action~\eqref{eq:RRaction} leads to the equations of motion
\begin{align}
d\star H=\sum_{k}\mu_{k}\delta_{\,9-k}(\mathcal{B}_{k}\hookrightarrow \mathcal{M}_{10})Y_{k}.
\end{align}
We have to remember however that in type-IIB theory~$C_{p+1}$ is dual to~$C_{9-p}$ 
($C_{4}$ is self-dual,~$H_{5}=\star H_{5}$). 
This means that~D$p$-branes are electric sources of~$C_{p+1}$,
whose magnetic sources are~D$(9-p)$-branes. With this in mind, the
Bianchi identity derived from the action~\eqref{eq:RRaction} reads
\begin{align}
dH=-\sum_{k}\mu_{k}\delta_{\,9-k}(\mathcal{B}_{k}\hookrightarrow \mathcal{M}_{10})\overline{Y}_{k},
\label{eq:bianchiH}
\end{align}
where~$\overline{Y}_{k}$ is obtained from~$Y_{k}$ by taking the complex conjugate of the
(fundamental) Chan-Paton bundle.
The conclusion is that the presence of the brane leads to a modification of the original field strength~$H=dC$, 
which now picks up terms depending on the brane gauge field strength and curvature
\begin{align}
H=dC-\sum_{k}\mu_{k}\delta_{\,9-k}(\mathcal{B}_{k}\hookrightarrow \mathcal{M}_{10})\overline{Y}^{0}_{k}.
\label{eq:H_inflow_modified}
\end{align}
Originally, the field~$C$ was gauge invariant. Requiring however 
that~\eqref{eq:H_inflow_modified} remains invariant leads to the transformation
\begin{align}
\delta C=\sum_{k}\mu_{k}\delta_{\,9-k}(\mathcal{B}_{k}\hookrightarrow \mathcal{M}_{10})\overline{Y}^{1}_{k},
\label{eq:transfC}
\end{align} 
where~$Y^{1}_{k}$ is defined by~$\delta Y^{0}_{k}=dY^{1}_{k}$ and similarly for~$\overline{Y}^{1}_{k}$. 
An important point here is that
the modification of the gauge transformation of the $C$~field is restricted to the brane worldvolumes.

We can compute now the gauge variation of the R-R action~\eqref{eq:RRaction}. Implementing the gauge transformation~\eqref{eq:transfC}
and the Bianchi identity~\eqref{eq:bianchiH}, we find
\begin{align}
\delta S_{\rm RR}&=-{1\over 2}\sum_{j,k}\mu_{j}\mu_{k}\int_{\mathcal{M}_{10}}
\delta_{\,9-j}(\mathcal{B}_{j}\hookrightarrow \mathcal{M}_{10})
\delta_{\,9-k}(\mathcal{B}_{k}\hookrightarrow \mathcal{M}_{10})
(Y_{j}\overline{Y}_{\!\!k})^{(1)},
\end{align}
where we have written
\begin{align}
(Y_{j}\overline{Y}_{\!\!k})^{(1)}\equiv N_{j}\overline{Y}_{\!\!k}^{1}+Y_{j}^{1}\overline{Y}_{k}.
\label{eq:YjYk(1)}
\end{align}
The product of the two delta forms can be reduced to a single one
taking into account the property~\cite{CY}
\begin{align}
\delta_{\,9-j}(\mathcal{B}_{j}\hookrightarrow \mathcal{M}_{10})&
\delta_{\,9-k}(\mathcal{B}_{k}\hookrightarrow \mathcal{M}_{10}) \nonumber \\[0.2cm]
&= 
\delta_{\,9-j}(\mathcal{B}_{j}\cap\mathcal{B}_{k}\hookrightarrow \mathcal{M}_{10})e(TN_{j}\cap TN_{k}).
\end{align}
With this, the anomalous variation of~$S_{\rm RR}$ can be expressed as
\begin{align}
\delta S_{\rm RR}&=-{1\over 2}\sum_{j,k}\mu_{j}\mu_{k}\int_{\mathcal{M}_{10}}
\delta_{\,9-j}(\mathcal{B}_{j}\cap\mathcal{B}_{k}\hookrightarrow \mathcal{M}_{10})e(TN_{j}\cap TN_{k})
(Y_{j}\overline{Y}_{\!\!k})^{(1)},
\label{eq:SRRgauge_var_final}
\end{align}
which, as we see, is supported on the D-brane intersections.  

The notation used in eq.~\eqref{eq:YjYk(1)} was not unintentional. The superscript on the 
left-hand side indicates that
this term is obtained by descend from the invariant
polynomial~$Y_{j}\overline{Y}_{\!\!k}$
\begin{align}
Y_{j}\overline{Y}_{\!\!k}&=N_{j}N_{k}+d\big(N_{j}\overline{Y}_{k}^{0}+N_{k}Y_{j}^{0}+
Y_{j}^{0}d\overline{Y}_{\!\!k}^{0}\big) \equiv N_{j}N_{k}+d(Y_{j}\overline{Y}_{\!\!k})^{(0)},
\nonumber \\[0.2cm]
\delta(Y_{j}\overline{Y}_{\!\!k})^{(0)}&=d\big(N_{j}\overline{Y}^{1}_{\!\!k}
+Y^{1}_{j}\overline{Y}_{\!\!k}\big)
 \equiv d(Y_{j}\overline{Y}_{\!\!k})^{(1)}.
\end{align}
This means that the gauge variation~\eqref{eq:SRRgauge_var_final} can be derived from the brane anomaly polynomial
\begin{align}
I'_{\rm bulk}&=-{1\over 4\pi}\sum_{j,k}\mu_{j}\mu_{k}
Y_{j}\overline{Y}_{\!\!k}e(TN_{j}\cap TN_{k}).
\label{eq:bulk_polynomial_braneinters}
\end{align}
where the factor of~$4\pi$ comes
from the global normalization of the anomaly.

After all these calculations, 
we can finally make contact with the D-brane anomaly polynomial 
in~\eqref{eq:anomaly_pol_brane}.
Particularizing the bulk anomaly 
polynomial~\eqref{eq:bulk_polynomial_braneinters} to a single stack of D$p$-branes
\begin{align}
I'_{\rm bulk}=-{\mu^{2}\over 4\pi} Y\overline{Y}e(TN),
\end{align} 
the condition that the anomalous variation of the bulk R-R action cancels
the anomaly on the brane determines the invariant polynomial~$Y$ to be
\begin{align}
Y=-{\sqrt{4\pi}\over \mu}\mbox{ch}(\mathcal{F})\sqrt{\widehat{A}(T\!\mathcal{B}_{p})\over \widehat{A}(TN)}.
\end{align}
This fixes the D$p$-brane R-R couplings in the action~\eqref{eq:RRaction_ansatz}
\begin{align}
S_{\rm int}=T_{p}\int_{\mathcal{B}_{p}}C\,\mbox{ch}(\ell_{s}^{2}\mathcal{F})
\sqrt{\widehat{A}(\ell_{s}^{2}\mathcal{R}_{T})\over \widehat{A}(\ell_{s}^{2}\mathcal{R}_{N})},
\label{eq:Dpbranecoupling}
\end{align}
where, 
for the sake of clarity, we have indicated the dependence of the $A$-roof genera on the curvatures
of the tangent and normal bundles. We have also restored the powers of the string length and the
D$p$-brane tension is defined as
\begin{align}
T_{p}\equiv {2\pi\over \ell_{s}^{p+1}}=2\pi(4\pi^{2}\alpha')^{-{p+1\over 2}}.
\end{align}
It should be pointed out that although our analysis has focused on the type-IIB theory, the
action~\eqref{eq:Dpbranecoupling} is valid as well in type-IIA string theory and even~$p$.
As an example, let us work out the case of a D3-brane. Extracting the terms of
rank four in the integrand, we have
\begin{align}
S_{\rm int}&=T_{3}
\int_{\mathcal{B}_{3}}\left[NC_{4}+{i\ell_{s}^{2}\over 2\pi}C_{2}{\rm tr}_{\rm f}\mathcal{F}
+{\ell_{s}^{4}\over 384\pi^{2}}C_{0}\Big(N\,{\rm Tr\,}\mathcal{R}^{2}-1440\,{\rm tr\,}\mathcal{F}^{2}\Big)\right].
\end{align}
where, to make things simpler, we have also set~$\mathcal{R}_{N}=0$. 

A similar analysis can be carried out for orientifold planes~\cite{MorScrSe,ScrSe}, 
where anomalies are originated in 
self-dual tensor fields. The resulting coupling  
is expressed using the Hirzebruch polynomial introduced in eq.~\eqref{eq:Isd_grav}
\begin{align}
S_{\rm int}=-2^{p-4}T_{p}\int_{\mathcal{O}_{p}} C\,\sqrt{L(\ell_{s}^{2}\mathcal{R}_{T}/4)\over L(\ell_{s}^{2}\mathcal{R}_{N}/4)},
\label{eq:Opcoupling}
\end{align}
where the square root is expanded in terms of the Pontrjagin classes for the tangent and normal bundle
\begin{align}
\sqrt{L(\ell_{s}^{2}\mathcal{R}_{T}/4)\over L(\ell_{s}^{2}\mathcal{R}_{N}/4)}&=
1+{\ell_{s}^{4}\over 96}\big(p_{1}-p'_{1}\big) \nonumber \\[0.2cm]
&+{\ell_{s}^{8}\over 92160}\Big(448 p_{2}-448 p_{2}'-91p_{1}^{2}+19 p'^{2}_{1}-10 p_{1}p'_{1}\Big)+\ldots
\end{align}
Notice that since there are no open strings attached to the orientifold plane, only gravitational 
anomalies may arise.

These anomalous couplings have an interesting
connection with the GS mechanism~\cite{MorScrSe}. In the modern language, type-I string theory 
is formulated as type-IIB theory in the presence of 32~D9-branes and one orientifold O9-plane. We can use  
expressions~\eqref{eq:Dpbranecoupling}
and~\eqref{eq:Opcoupling} to compute how 
the R-R two-form~$C_{2}$ and its dual~$C_{6}$ couple to these space-filling defects. Adding their
corresponding contributions and keeping the piece of rank 10, we find the relevant terms in the action
to be\footnote{Interestingly, the 
tadpole term proportional to~$C_{10}$ cancels as a consequence of the gauge group being~SO($32$).}
\begin{align}
S_{\rm int}&\supset {2\pi\over \ell_{s}^{6}}\int_{\mathcal{M}_{10}}
\bigg[C_{6}\big(\mbox{ch}_{\rm f,2}-p_{1}\big) \nonumber \\[0.2cm]
&\left.+\ell_{s}^{4}C_{2}\left(\mbox{ch}_{\rm f,4}-{1\over 48}p_{1}\mbox{ch}_{\rm f,2}+{1\over 64}p_{1}^{2}
-{1\over 48}p_{2}\right)\right],
\label{eq:C6C2actionD9O9}
\end{align}
where in the Chern characters we indicated explicitly that traces are computed in the fundamental
of~SO($32$). The coupling of~$C_{6}$ determines the Bianchi identity for~$C_{2}$ 
to be $dH_{3}=\ell_{s}^{2}(\mbox{ch}_{\rm f,2}-p_{1})$, imposing the modification of the field strength
$H_{3}=dC_{2}+\ell_{s}^{2}(\omega_{3}^{0}-\Omega_{3}^{0})$. 
This reproduces
eq.~\eqref{eq:new_field_strength_B} with~$B=C_{2}$. 
Moreover,  
writing the Chern characters explicitly in terms of traces, we recover as well the GS counterterm~\eqref{eq:X8_general10D_traces} from the second term on the right-hand side
of~\eqref{eq:C6C2actionD9O9}.
This shows that the GS mechanism can be regarded
as an instance of anomaly inflow, where the anomaly in the open string sector is cancelled by 
gauge and gravitational charge influx provided by the closed string R-R rank-two antisymmetric tensor field,
which now has gauge charge due to its modified field strength.

\section{A modern take on anomalies}

Our overview of anomalies has been restricted to those jeopardizing the invariance of 
quantum theories under {\em infinitesimal} diffeomorphisms and gauge transformations, which 
for this very reason are visible 
in perturbation theory. But anomalies can also affect transformations not in the connected component
of the identity. One example is
Witten's global anomaly~\cite{Witten_anom}, which destroys the consistency of 
four-dimensional SU(2) gauge theories with an odd number of left-handed fundamental fermions. 
As for the role played by anomaly cancellation in string theory, we focused on {\em spacetime} 
(i.e., target space) anomalies,
although by no means they are the only type constraining consistent string models.
Indeed, the cancellation of the worldsheet conformal anomaly is the crucial element
leading to the notion of critical dimension, while modular invariance ensures the absence 
of global gravitational anomalies on the worldsheet.

\begin{figure}[t]
\centerline{\includegraphics[scale=0.3]{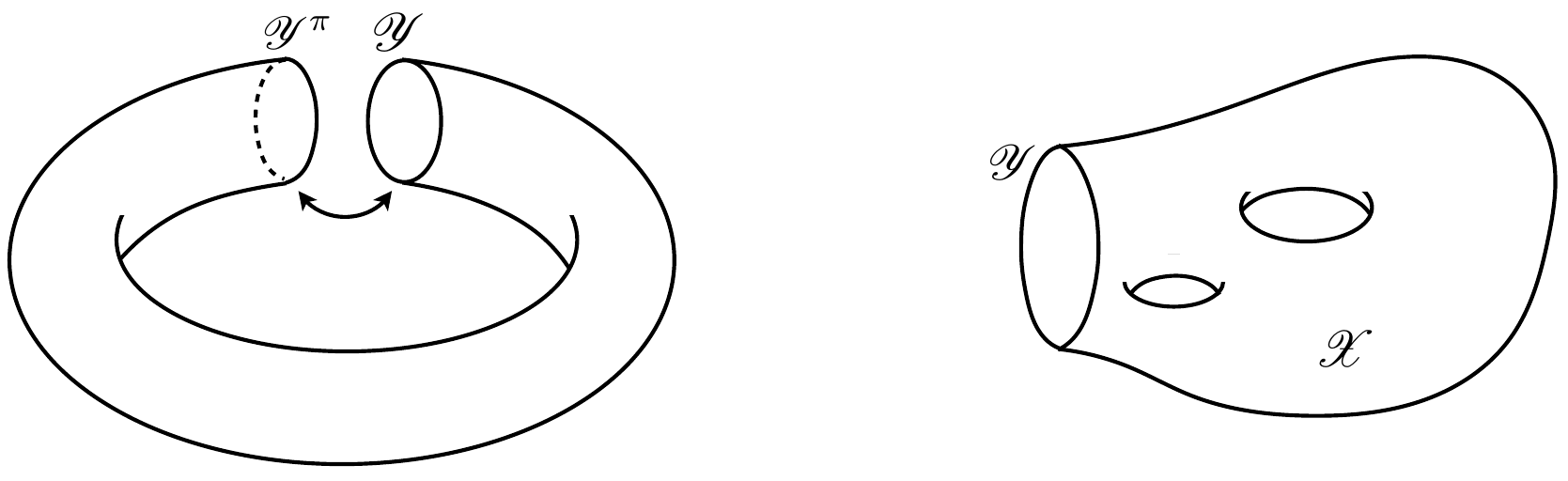}}
\caption[]{{\bf Left:} construction of the mapping torus associated with a global transformation~$\pi$. The two boundaries of the 
cylinder~$\mathcal{Y}\times[0,1]$ are glued together modulo~$\pi$. {\bf Right:} the Euclidean spacetime~$\mathcal{Y}$ is seen 
as the boundary of a~$(D+1)$-dimensional manifolds with all the relevant mathematical structures on~$\mathcal{Y}$ properly 
extended.}
\label{fig:mapping}
\end{figure}

In page~\pageref{page:anomaly_polynomial} we learned how perturbative anomalies on an even $D$-dimensional
manifold~$\mathcal{Y}$ are obtained from the variation of a~$(D+1)$-dimensional Chern-Simons form,
related to the index of a Dirac operator in~$D+2$ dimensions. 
A similar structure exists for global anomalies~\cite{Witten_global_grav} where the
central role is played by the $\eta$-invariant, defined as the regularized sum of the signs of the 
eigenvalues of the Dirac operator
\begin{align}
\eta\equiv \Big[\sum_{\lambda_{i}\neq 0}\mbox{sign}(\lambda_{i})\Big]_{\rm reg}
=\lim_{\epsilon\rightarrow 0^{+}}
\sum_{\lambda_{i}\neq 0}{\lambda_{i}\over |\lambda_{i}|}e^{-\epsilon|\lambda_{i}|}.
\end{align}
The variation of the effective action under a {\em global} transformation~$\pi$ 
is given in terms of the $\eta$-invariant of the associated 
mapping torus~$\mathcal{T}_{\pi}\equiv (\mathcal{Y}\times S^{1})_{\pi}$. This is 
constructed from the cylinder~$\mathcal{Y}\times [0,1]$ by identifying its boundaries
modulo the transformation~$\pi$, as shown on the left of 
fig.~\ref{fig:mapping}. 
For a Weyl fermion, the result is
\begin{align}
\Delta_{\pi}\Gamma_{\rm eff}\equiv \Gamma_{\rm eff}^{\pi}-\Gamma_{\rm eff}={i\pi\over 2}\eta_{\mathcal{T}_{\pi}}.
\label{eq:global_anomaly_weyl}
\end{align}
The
Atiyal-Patodi-Singer (APS) theorem relates this $\eta$-invariant to the index of
the Dirac operator on a~$(D+2)$-dimensional manifold~$\mathcal{B}$ such 
that~$\partial\mathcal{B}=\mathcal{T}_{\pi}$
\begin{align}
{1\over 2}\eta_{\mathcal{T}_{\pi}}=\mbox{index}_{\mathcal{B}}(iD\!\!\!\!\!/)-\int_{\mathcal{B}}
[\widehat{A}(\mathcal{R})\mbox{ch}(\mathcal{F})]_{D+2},
\end{align}
where the second term on the right-hand side vanishes in the absence of perturbative anomalies.

The APS~$\eta$-invariant is in fact the key to a whole new approach to 
anomalies building on the notion of anomaly inflow and 
with deep mathematical implications~\cite{dai_freed,df_cm,df_hep}. 
Let us take our Euclidean $D$-dimensional
manifold of interest~$\mathcal{Y}$ to be
the boundary of some manifold~$\mathcal{X}$ (see the right of fig.~\ref{fig:mapping}). 
In order to define fermions properly,
we require the spin/pin structure on~$\mathcal{Y}$ to be smoothly extended to~$\mathcal{X}$. Moreover, 
boundary conditions for fermions on~$\mathcal{Y}=\partial\mathcal{X}$ have to be carefully chosen so
the Dirac operator on~$\mathcal{X}$ is self-adjoint (see~\cite{dai_freed,df_cm,df_hep} for 
the technical details).
Once all mathematical subtleties are properly handled, the partition function
for a Weyl fermion on~$\mathcal{Y}$ can be written as
\begin{align}
Z\equiv e^{-\Gamma_{\rm eff}}=|\mbox{Pf}(iD\!\!\!\!\!/\,)|
\exp\left(-{i\pi\over 2}\eta_{\mathcal{X}}\right)\exp\left(-2\pi i\int_{\mathcal{X}}I^{0}_{D+1}\right).
\label{eq:partfunc_Dai_Freed}
\end{align}
This expression 
is independent on changes on either the metric or the gauge field on~$\mathcal{X}$ 
as far as these
do not affect their values on~$\mathcal{Y}$. This means that, despite appearances, it only 
depends on field theory data on the boundary. 

\begin{figure}[t]
\centerline{\includegraphics[scale=0.35]{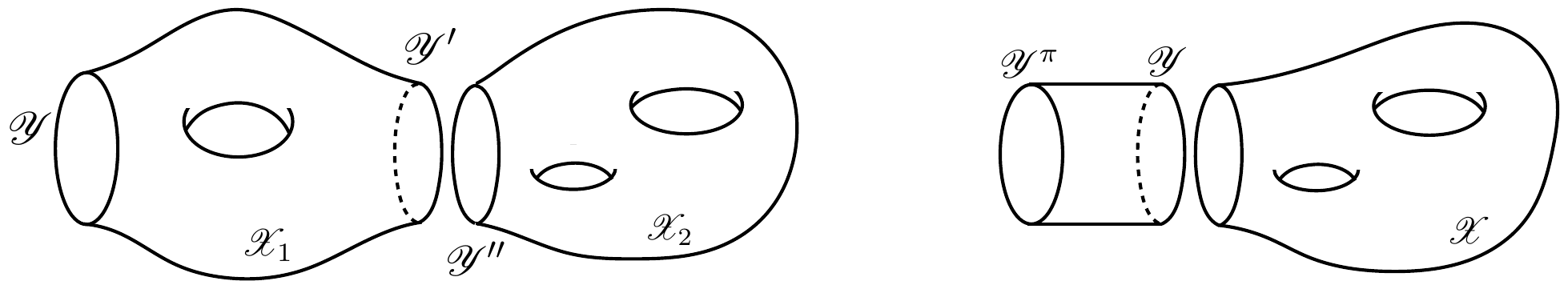}}
\caption[]{{\bf Left:} gluing two surfaces together 
along their boundaries~$\mathcal{Y}'$ and~$\mathcal{Y}''$. {\bf Right:} global transformations
are implemented by
attaching to~$\mathcal{Y}=\partial\mathcal{X}$ the appropriate mapping torus.}
\label{fig:cobordisms}
\end{figure}

The fermion partition function~\eqref{eq:partfunc_Dai_Freed} smoothly connects with the standard analysis of
perturbative and global anomalies. For the first type, the $\eta$-invariant remains unchanged while the variation of the Chern-Simons form
gives the known expression for the integrated anomaly\footnote{In Euclidean space
the anomaly only affects the imaginary part of the effective action, 
so~$|\mbox{Pf}(iD\!\!\!\!\!/\,)|$ remains invariant.}. 
The partition function may also
change under transformations not connected with the identity, and here is where the sewing properties of the~$\eta$-invariant come in handy. 
Consider two manifolds~$\mathcal{X}_{1}$ and~$\mathcal{X}_{2}$ being glued together by their boundaries, as 
shown on the left
of fig.~\ref{fig:cobordisms}. The~$\eta$-invariant associated with the glued manifold~$\mathcal{X}_{1}\sqcup\mathcal{X}_{2}$ is given by
\begin{align}
e^{-{i\pi\over 2}\eta_{\mathcal{X}_{1}\sqcup\mathcal{X}_{2}}}=e^{-{i\pi\over 2}\eta_{\mathcal{X}_{1}}}
e^{-{i\pi\over 2}\eta_{\mathcal{X}_{2}}}.
\label{eq:cobordism_eta_inv}
\end{align}
We can apply this relation to compute the change in the partition function~\eqref{eq:partfunc_Dai_Freed} under a global
transformation~$\pi$, implemented by gluing the corresponding mapping torus to the boundary
of~$\mathcal{X}$ (see the right of fig.~\ref{fig:cobordisms}). 
Using eq.~\eqref{eq:cobordism_eta_inv}, we find
\begin{align}
{Z^{\pi}\over Z}=e^{-{i\pi\over 2}\eta_{\mathcal{T}_{\pi}}},
\end{align}
which reproduces the form of the 
global anomaly given in~\eqref{eq:global_anomaly_weyl}. In the previous discussion 
we assumed tacitly that~$D$ is even and focused our attention on gauge and gravitational anomalies. 
For~$D$ odd, the Chern-Simons term in the partition function is absent and the standard treatment of parity anomaly in terms of
the $\eta$-invariant~\cite{AGdPM} is retrieved.

The phase in eq.~\eqref{eq:partfunc_Dai_Freed} can be interpreted as the partition function of an ``anomaly'' topological field theory
defined on~$\mathcal{X}$ that includes fermions. An important question 
is whether there are ambiguities associated with the
choice of the manifold~$\mathcal{X}$, the so-called
Dai-Freed anomalies. In fact, the consistency of the theory on~$\mathcal{Y}$ 
requires then that the 
fermion partition function does not depend on the higher-dimensional manifold~$\mathcal{X}$ used and therefore that it is
free from Dai-Freed anomalies.
Defining the partition function using 
two different manifolds~$\mathcal{X}_{1}$ and~$\mathcal{X}_{2}$ with~$\mathcal{Y}
=\partial\mathcal{X}_{1}=\partial\mathcal{X}_{2}$, this condition takes the form
\begin{align}
1={Z_{1}\over Z_{2}}=e^{{i\pi\over 2}\eta_{\mathcal{X}_{1}\sqcup\overline{\mathcal{X}}_{2}}},
\label{eq:1=Z1/Z2}
\end{align}
where~$\mathcal{X}_{1}\sqcup\overline{\mathcal{X}}_{2}$ is the closed manifold resulting from gluing~$\mathcal{X}_{1}$ 
and~$\mathcal{X}_{2}$ along their common 
boundary and the bar indicates orientation reversal (the  
property~$\eta_{\overline{\mathcal{X}}}=-\eta_{\mathcal{X}}$ was also used). Notice 
that~$\mathcal{X}_{1}\sqcup\overline{\mathcal{X}}_{2}$ can be seen as the boundary of a~$(D+2)$-dimensional manifold. 
Using the Stokes theorem together with~$dI^{0}_{D+1}=I_{D+2}$, we find that
the integral of the Chern-Simons term 
over the closed manifold gives an integer resulting in a trivial phase.

Equation~\eqref{eq:1=Z1/Z2} means that the 
cancellation of Dai-Freed anomalies is equivalent to the requirement that
the higher-dimensional anomaly topological field theory is trivial. This condition
comprises the absence of standard gauge and gravitational 
(perturbative and global) anomalies, but in fact goes much beyond. It includes a wider class of consistency 
conditions to be imposed on quantum field theories. This has found interesting application
in string theory~\cite{df_st}, high-energy phenomenology~\cite{df_hep}, 
and the physics of the topological phases of matter~\cite{df_cm}.

But there are more interesting mathematics lurking behind all this.
Two closed $D$-dimensional manifolds~$\mathcal{Y}_{1}$ and~$\mathcal{Y}_{2}$ 
are in the same equivalence 
bordism class if their disjoint union~$\mathcal{Y}_{1}\sqcup\overline{\mathcal{Y}}_{2}$
is the boundary of some~$(D+1)$-dimensional manifold. An example of this 
is illustrated on the left picture of fig.~\ref{fig:cobordisms},
where from~$\mathcal{Y}\sqcup\mathcal{Y}'=\partial\mathcal{X}_{1}$ we see that~$\mathcal{Y}$ and~$\overline{\mathcal{Y}}'$ belong to the 
same bordism class. The discussion above indicates that the
cancellation of Dai-Freed anomalies means that 
consistent fermion theories are invariant under bordisms. In other words, it does not matter what representative~$\mathcal{X}$ within a
bordism equivalence class we choose to define the theory.

This bordism invariance has led to a fascinating 
connection with category theory, one of the booming topics in contemporary mathematics
(see~\cite{categories_rev} for a physicist-oriented review). 
To discuss just the rudiments of category theory and how they apply to 
the analysis of quantum field theory anomalies lies
way beyond the scope of this brief overview. The reader can find a
nice introduction to this topic in ref.~\cite{monnier}.  
It is in any case interesting how, since their diagrammatic inception in~1969~\cite{discovery},
the subject of quantum field theory 
anomalies has provided a fertile ground for the application of new mathematics. 
It was from the late 1970s on~\cite{AS}
that the beautiful and insightful connection with the index theorems emerged. This did not just 
clarify many features of anomalies already identified
in the diagrammatic approach, such as their saturation at one loop, but also highlighted its
very general nature independent of the technical details of the Feynman diagram 
computations where they were first identified.
Although the categorial approach to anomalies
is still very much under exploration, the expectation exists that it
could lead to a deepening of our understanding of
quantum fields in some way comparable to
what was achieved by the implementation of differential geometry techniques.

\section*{Acknowledgments} 

We thank Juan L. Ma\~nes and Manuel Valle for valuable discussions on topics related to the subject of
this review. M.A.V.-M. acknowledges financial support from the
Spanish Science Ministry through research grants PGC2018-094626-B-C22 and PID2021-123703NB-C22
(MCIU/AEI/FEDER, EU), as well as from Basque Government grant
IT1628-22.


\begin{thebibliography}{99}

\bibitem{QFT_book}
L.~\'Alvarez-Gaum\'e, M.~\'A.~V\'azquez-Mozo, \href{https://doi.org/10.1007/978-3-642-23728-7}{\emph{{An Invitation to Quantum Field Theory}}},
Springer, 2012.

\bibitem{anomalies_rev}
L.~\'Alvarez-Gaum\'e, 
\href{https://doi.org/10.1007/978-1-4757-0363-4_4}{\emph{{An Introduction to Anomalies}}},  in
  \emph{{``\href{https://doi.org/10.1007/978-1-4757-0363-4}{Fundamental 
  Problems of Gauge Field Theory}''}}, {Plenum Press},
  1985.
\\
R.~A.~Bertlmann, \href{https://doi.org/10.1093/acprof:oso/9780198507628.001.0001}{
\emph{{Anomalies in Quantum Field Theory}}}, Oxford University
  Press, 1996.
\\
K.~Fujikawa and H.~Suzuki, \href{https://doi.org/10.1093/acprof:oso/9780198529132.001.0001}{\emph{{Path 
Integrals and Quantum Anomalies}}}, Oxford University Press, 2004.
\\
A.~Bilal,
{\it Lectures on Anomalies,}
\href{https://doi.org/10.48550/arXiv.0802.0634}{arXiv:0802.0634 [hep-th]}.
  
\bibitem{harvey_review}
J.~A.~Harvey,
{\it TASI Lectures on Anomalies,}
\href{https://doi.org/10.48550/arXiv.hep-th/0509097}{arXiv:hep-th/0509097 [hep-th]}.

\bibitem{BZ_NPB}
W.~A.~Bardeen and B.~Zumino,
{\it Consistent and Covariant Anomalies in Gauge and Gravitational Theories,}
\href{https://doi.org/10.1016/0550-3213(84)90322-5}{Nucl. Phys. B \textbf{244} (1984) 421}.

\bibitem{AG_GinspargAP}
L.~\'Alvarez-Gaum\'e and P.~Ginsparg,
{\it The Structure of Gauge and Gravitational Anomalies,}
\href{https://doi.org/10.1016/0003-4916(85)90087-9}{Annals Phys. \textbf{161} (1985) 423}.

\bibitem{nakahara}
M.~Nakahara, \emph{Geometry, Topology and Physics (2nd edition)}, Taylor \& Francis, 2003.

\bibitem{zumino}
B. Zumino, 
{\it Chiral anomalies and differential geometry}, 
in ``Relativity, groups and topology'', Elsevier (1983).
\\
M.~F.~Atiyah and I.~M.~Singer,
{\it Dirac Operators Coupled to Vector Potentials,}
\href{https://doi.org/10.1073/pnas.81.8.2597}{Proc. Nat. Acad. Sci. \textbf{81} (1984) 2597}.
\\
B.~Zumino, Y.~S.~Wu and A.~Zee,
{\it Chiral Anomalies, Higher Dimensions, and Differential Geometry,}
\href{https://doi.org/10.1016/0550-3213(84)90259-1}{Nucl. Phys. B \textbf{239} (1984) 477}.

\bibitem{AG_GinspargNPB}
L.~\'Alvarez-Gaum\'e and P.~Ginsparg,
\emph{The Topological Meaning of Nonabelian Anomalies},
\href{https://doi.org/10.1016/0550-3213(84)90487-5}{Nucl. Phys. B \textbf{243} (1984) 449}.

\bibitem{AG_Witten}
L.~\'Alvarez-Gaum\'e and E.~Witten,
{\it Gravitational Anomalies},
\href{https://doi.org/10.1016/0550-3213(84)90066-X}{Nucl. Phys. B \textbf{234} (1984) 269}.

\bibitem{GS1}
M.~B.~Green and J.~H.~Schwarz,
{\it Anomaly Cancellation in Supersymmetric D=10 Gauge Theory and Superstring Theory,}
\href{https://doi.org/10.1016/0370-2693(84)91565-X}{Phys. Lett. B \textbf{149} (1984) 117}.

\bibitem{GS2}
M.~B.~Green and J.~H.~Schwarz,
{\it The Hexagon Gauge Anomaly in Type I Superstring Theory,}
\href{https://doi.org/10.1016/0550-3213(85)90130-0}{Nucl. Phys. B \textbf{255} (1985) 93}.

\bibitem{string_reviews}
L.~\'Alvarez-Gaum\'e and M.~\'A.~V\'azquez-Mozo, {\it Topics in String Theory and Quantum 
Gravity}, in: ``Gravitation and Quantizations'', Proceedings of the 1992 Les Houches 
Summer School, Elsevier 1995 [\href{https://doi.org/10.48550/arXiv.hep-th/9212006}{hep-th/9212006}].\\
J.~Polchinski, {\it String Theory vols. 
\href{https://doi.org/10.1017/CBO9780511816079}{I} \& 
\href{https://doi.org/10.1017/CBO9780511618123}{II}}, Cambridge 1998. \\
K.~Becker, M.~Becker and J.~H.~Schwarz, 
\href{https://doi.org/10.1017/CBO9780511816086}{\it String Theory and M-Theory: A Modern Introduction}, 
Cambridge 2006. \\
L.~E.~Ib\'a\~nez and \'A.~M.~Uranga, 
\href{https://doi.org/10.1017/CBO9781139018951}{\it String Theory and Particle Physics: An Introduction
to String Phenomenology}, Cambridge 2012. \\
E.~Kiritsis, {\it String Theory in a Nutshell}, Princeton 2019.

\bibitem{johnson}
C.~V.~Johnson, \href{https://doi.org/10.1017/CBO9780511606540}{\it D-branes}, Cambridge 2003.

\bibitem{GSW}
M.~B.~Green, J. H.~Schwarz and E.~Witten, {\it Superstring Theory vols 
\href{https://doi.org/10.1017/CBO9781139248563}{I} \& 
\href{https://doi.org/10.1017/CBO9781139248570}{II}}, Cambridge 1987.

\bibitem{CaiPol}
J.~Polchinski and Y.~Cai,
{\it Consistency of Open Superstring Theories,}
\href{https://doi.org/10.1016/0550-3213(88)90382-3}{Nucl. Phys. B \textbf{296} (1988) 91}.

\bibitem{GS_finitude}
M.~B.~Green and J.~H.~Schwarz,
{\it Infinity Cancellations in SO(32) Superstring Theory,}
\href{https://doi.org/10.1016/0370-2693(85)90816-0}{Phys. Lett. B \textbf{151} (1985) 21}.

\bibitem{BdRdWvN}
A.~H.~Chamseddine,
{\it Interacting Supergravity in Ten-Dimensions: The Role of the Six-Index Gauge Field,}
\href{https://doi.org/10.1103/PhysRevD.24.3065}{Phys. Rev. D \textbf{24} (1981) 3065}.
\\
E.~Bergshoeff, M.~de Roo, B.~de Wit and P.~van Nieuwenhuizen,
{\it Ten-Dimensional Maxwell-Einstein Supergravity, Its Currents, and the Issue of Its Auxiliary Fields,}
\href{https://doi.org/10.1016/0550-3213(82)90050-5}{Nucl. Phys. B \textbf{195}
(1982) 97}.

\bibitem{chapline_manton}
G.~F.~Chapline and N.~S.~Manton,
{\it Unification of Yang-Mills Theory and Supergravity in Ten-Dimensions,}
\href{https://doi.org/10.1016/0370-2693(83)90633-0}{Phys. Lett. B \textbf{120} 
(1983) 105}.

\bibitem{WZ_lagrangian}
J.~Wess and B.~Zumino,
{\it Consequences of anomalous Ward identities,}
\href{https://doi.org/10.1016/0370-2693(71)90582-X}{Phys. Lett. B \textbf{37} 
(1971) 95}.

\bibitem{open_gen_anom_anal}
M.~Hayashi, N.~Kawamoto, T.~Kuramoto and K.~Shigemoto,
{\it Gravitational Anomaly Cancellation in Type I Superstring Theory,}
\href{https://doi.org/10.1016/0550-3213(88)90677-3}{Nucl. Phys. B \textbf{296} 
(1988) 373}.

\bibitem{princeton_quartet}
D.~J.~Gross, J.~A.~Harvey, E.~J.~Martinec and R.~Rohm,
{\it Heterotic String Theory. 1. The Free Heterotic String,}
\href{https://doi.org/10.1016/0550-3213(85)90394-3}{Nucl. Phys. B \textbf{256} (1985) 253}.
\\
D.~J.~Gross, J.~A.~Harvey, E.~J.~Martinec and R.~Rohm,
{\it Heterotic String Theory. 2. The Interacting Heterotic String,}
\href{https://doi.org/10.1016/0550-3213(86)90146-X}{Nucl. Phys. B \textbf{267} (1986) 75}.

\bibitem{type-II_cancellation}
M.~Hayashi, N.~Kawamoto, T.~Kuramoto and K.~Shigemoto,
{\it Modular Invariance and Gravitational Anomaly in Type {II} Superstring Theory,}
\href{https://doi.org/10.1016/0550-3213(87)90592-X}{Nucl. Phys. B \textbf{294} 
(1987) 459}.

\bibitem{heterotic_anomalies}
A.~N.~Schellekens and N.~P.~Warner,
{\it Anomalies and Modular Invariance in String Theory,}
\href{https://doi.org/10.1016/0370-2693(86)90760-4}{Phys. Lett. B \textbf{177} 
(1986) 317}.
\\
H.~Suzuki and A.~Sugamoto,
{\it Role of Modular Invariance in Evaluation of Gauge and Gravitational Anomalies in the Heterotic String,}
\href{https://journals.aps.org/prl/abstract/10.1103/PhysRevLett.57.1665}{Phys. Rev. Lett. \textbf{57} 
(1986) 1665}.
\\
W.~Lerche, B.~E.~W.~Nilsson and A.~N.~Schellekens,
{\it Heterotic String-Loop Calculation of the Anomaly Cancelling Term,}
\href{https://doi.org/10.1016/0550-3213(87)90397-X}{Nucl. Phys. B \textbf{289} 
(1987) 609}.
\\
D.~J.~Gross and P.~F.~Mende,
{\it Modular Subgroups, Odd Spin Structures and Gauge Invariance in the Heterotic String,}
\href{https://doi.org/10.1016/0550-3213(87)90489-5}{Nucl. Phys. B \textbf{291} 
(1987) 653}.
\\
D.~Kutasov,
{\it Modular Invariance, Chiral Anomalies and Contact Terms,}
\href{https://doi.org/10.1016/0550-3213(88)90330-6}{Nucl. Phys. B \textbf{307} 
(1988) 417}.
\\
W.~Lerche, A.~N.~Schellekens and N.~P.~Warner,
{\it Lattices and Strings,}
\href{https://doi.org/10.1016/0370-1573(89)90077-X}{Phys. Rept. \textbf{177} 
(1989) 1}.

\bibitem{SO(16)}
L.~\'Alvarez-Gaum\'e, P.~H.~Ginsparg, G.~W.~Moore and C.~Vafa,
{\it An $\mathrm{O}(16)\times \mathrm{O}(16)$ Heterotic String},
\href{https://doi.org/10.1016/0370-2693(86)91524-8}{Phys. Lett. B \textbf{171} (1986) 155}.

\bibitem{sugimoto}
S.~Sugimoto,
{\it Anomaly cancellations in type I D-9-anti-D-9 system and the USp(32) string theory,}
\href{https://doi.org/10.1143/PTP.102.685}{Prog. Theor. Phys. \textbf{102} (1999) 685}
[\href{https://doi.org/10.48550/arXiv.hep-th/9905159}{arXiv:hep-th/9905159 [hep-th]}].

\bibitem{bonnefoy_dudas}
Q.~Bonnefoy and E.~Dudas,
{\it Axions and anomalous U(1)\textquoteright{}s,}
\href{https://doi.org/10.1142/S0217751X1845001X}{Int. J. Mod. Phys. A \textbf{33} 
(2018) 1845001}
[\href{https://doi.org/10.48550/arXiv.1809.08256}{arXiv:1809.08256 [hep-ph]}].

\bibitem{MSZ}
J.~Ma\~nes, R.~Stora and B.~Zumino,
{\em Algebraic Study of Chiral Anomalies,}
\href{https://doi.org/10.1007/BF01208825}{Commun. Math. Phys. \textbf{102} (1985) 157}.

\bibitem{MMVVM18}
J.~L.~Ma\~nes, E.~Meg\'\i{}as, M.~Valle and M.~\'A.~V\'azquez-Mozo,
{\it Non-Abelian Anomalous (Super)Fluids in Thermal Equilibrium from Differential Geometry,}
\href{https://doi.org/10.1007/JHEP11(2018)076}{JHEP \textbf{11} (2018) 076}
[\href{https://doi.org/10.48550/arXiv.1806.07647}{arXiv:1806.07647 [hep-th]}].

\bibitem{CH}
C.~G.~Callan and J.~A.~Harvey,
{\it Anomalies and Fermion Zero Modes on Strings and Domain Walls,}
\href{https://doi.org/10.1016/0550-3213(85)90489-4}{Nucl. Phys. B \textbf{250} (1985) 427}.

\bibitem{BH}
J.~D.~Blum and J.~A.~Harvey,
{\it Anomaly inflow for gauge defects,}
\href{https://doi.org/10.1016/0550-3213(94)90580-0}{Nucl. Phys. B \textbf{416}
(1994) 119}
[\href{https://doi.org/10.48550/arXiv.hep-th/9310035}{arXiv:hep-th/9310035 [hep-th]}].

\bibitem{GW}
J.~Goldstone and F.~Wilczek,
{\it Fractional Quantum Numbers on Solitons,}
\href{https://doi.org/10.1103/PhysRevLett.47.986}{Phys. Rev. Lett. \textbf{47} 
(1981) 986}.

\bibitem{naculich}
S.~G.~Naculich,
{\it Axionic Strings: Covariant Anomalies and Bosonization of Chiral Zero Modes,}
\href{https://doi.org/10.1016/0550-3213(88)90400-2}{Nucl. Phys. B \textbf{296} 
(1988) 837}.

\bibitem{dbranes_polchinski}
J.~Polchinski,
{\it Dirichlet Branes and Ramond-Ramond charges,}
\href{https://doi.org/10.1103/PhysRevLett.75.4724}{Phys. Rev. Lett. \textbf{75} 
(1995) 4724}
[\href{https://doi.org/10.48550/arXiv.hep-th/9510017}{arXiv:hep-th/9510017 [hep-th]}].

\bibitem{DKL}
M.~J.~Duff, R.~R.~Khuri and J.~X.~Lu,
{\it String solitons,}
\href{https://doi.org/10.1016/0370-1573(95)00002-X}{Phys. Rept. \textbf{259} (1995) 213}
[\href{https://doi.org/10.48550/arXiv.hep-th/9412184}{arXiv:hep-th/9412184 [hep-th]}].

\bibitem{GHM}
M.~B.~Green, J.~A.~Harvey and G.~W.~Moore,
{\it I-brane inflow and anomalous couplings on d-branes,}
\href{https://doi.org/10.1088/0264-9381/14/1/008}{Class. Quant. Grav. \textbf{14} (1997) 47}
[\href{https://doi.org/10.48550/arXiv.hep-th/9605033}{arXiv:hep-th/9605033 [hep-th]}].

\bibitem{CY}
Y.~K.~E.~Cheung and Z.~Yin,
{\it Anomalies, branes, and currents,}
\href{https://doi.org/10.1016/S0550-3213(98)00115-1}{Nucl. Phys. B \textbf{517} (1998) 69}
[\href{https://doi.org/10.48550/arXiv.hep-th/9710206}{arXiv:hep-th/9710206 [hep-th]}].


\bibitem{szabo}
R.~J.~Szabo, 
\href{https://doi.org/10.1142/p741}{\it An Introduction to String Theory and D-brane 
Dynamics (2nd edition)}, Imperial College Press 2011.

\bibitem{seiberg_rev}
N.~Seiberg,
{\it Notes on theories with 16 supercharges,}
\href{https://doi.org/10.1016/S0920-5632(98)00128-5}{Nucl. Phys. B Proc. Suppl. \textbf{67} (1998) 158}
[\href{https://doi.org/10.48550/arXiv.hep-th/9705117}{arXiv:hep-th/9705117 [hep-th]}].

\bibitem{nash}
C.~Nash, {\it Topology and Quantum Field Theory}, Academic Press 1991.

\bibitem{MorScrSe}
J.~F.~Morales, C.~A.~Scrucca and M.~Serone,
{\it Anomalous couplings for D-branes and O-planes,}
\href{https://doi.org/10.1016/S0550-3213(99)00217-5}{Nucl. Phys. B \textbf{552} (1999) 291}
[\href{https://doi.org/10.48550/arXiv.hep-th/9812071}{arXiv:hep-th/9812071 [hep-th]}].

\bibitem{ScrSe}
C.~A.~Scrucca and M.~Serone,
{\it Anomalies and inflow on D-branes and O-planes,}
\href{https://doi.org/10.1016/S0550-3213(99)00357-0}{Nucl. Phys. B \textbf{556} (1999) 197}
[\href{https://doi.org/10.48550/arXiv.hep-th/9903145}{arXiv:hep-th/9903145 [hep-th]}].

\bibitem{Witten_anom}
E.~Witten,
{\it An SU(2) Anomaly,}
\href{https://doi.org/10.1016/0370-2693(82)90728-6}{Phys. Lett. B \textbf{117} (1982) 324}.

\bibitem{Witten_global_grav}
E.~Witten,
{\it Global Gravitational Anomalies},
\href{https://doi.org/10.1016/0370-2693(82)90728-6}{Commun. Math. Phys. \textbf{100} (1985) 197}.

\bibitem{dai_freed}
X.~Dai and D.~S.~Freed,
{\it $\eta$ invariants and determinant lines,}
\href{https://doi.org/10.1063/1.530747
}{J. Math. Phys. \textbf{35} (1994) 5155}
[\href{https://doi.org/10.48550/arXiv.hep-th/9405012}{arXiv:hep-th/9405012 [hep-th]}].
\\
E.~Witten and K.~Yonekura,
{\it Anomaly Inflow and the $\eta$-Invariant,} 
in: ``\href{https://doi.org/10.1142/12146}{Memorial Volume for Shoucheng Zhang}'', World Scientific 2021
[\href{https://doi.org/10.48550/arXiv.1909.08775}{arXiv:1909.08775 [hep-th]}].

\bibitem{df_cm}
E.~Witten,
{\it Fermion path integrals and topological phases,}
\href{https://doi.org/10.1103/RevModPhys.88.035001}{Rev. Mod. Phys. \textbf{88} (2016) 035001}
[\href{https://doi.org/10.48550/arXiv.1508.04715}{arXiv:1508.04715 [cond-mat.mes-hall]}].
\\
E.~Witten,
{\it The ''parity'' anomaly on an unorientable manifold,}
\href{https://doi.org/10.1103/PhysRevB.94.195150}{Phys. Rev. B \textbf{94} (2016) 195150}
[\href{https://doi.org/10.48550/arXiv.1605.02391}{arXiv:1605.02391 [hep-th]}].
\\
K.~Yonekura,
{\it Dai-Freed theorem and topological phases of matter,}
\href{https://doi.org/10.1007/JHEP09(2016)022}{JHEP \textbf{09} (2016), 022}
[\href{https://doi.org/10.48550/arXiv.1607.01873}{arXiv:1607.01873 [hep-th]}].

\bibitem{df_hep}
I.~Garc\'\i{}a-Etxebarria and M.~Montero,
{\it Dai-Freed anomalies in particle physics,}
\href{https://doi.org/10.1007/JHEP08(2019)003}{JHEP \textbf{08} (2019) 003}
[\href{https://doi.org/10.48550/arXiv.1808.00009}{arXiv:1808.00009 [hep-th]}].

\bibitem{AGdPM}
L.~\'Alvarez-Gaum\'e, S.~Della Pietra and G.~Moore,
{\it Anomalies and Odd Dimensions,}
\href{https://doi.org/10.1016/0003-4916(85)90383-5}{Annals Phys. \textbf{163} (1985) 288}.

\bibitem{df_st}
D.~S.~Freed and E.~Witten,
{\it Anomalies in string theory with D-branes,}
\href{https://dx.doi.org/10.4310/AJM.1999.v3.n4.a6}{Asian J. Math. \textbf{3} (1999) 819}
[\href{https://doi.org/10.48550/arXiv.hep-th/9907189}{arXiv:hep-th/9907189 [hep-th]}].

\bibitem{categories_rev}
B.~Coecke,~\'E.~O.~Paquette, 
\href{https://doi.org/10.1007/978-3-642-12821-9_3}{\it Categories for the Practicing Physicist}, in: 
``\href{https://doi.org/10.1007/978-3-642-12821-9}{New Structures in 
Physics}'' ed. B.~Coecke, Springer 2011.

\bibitem{monnier}
S.~Monnier,
{\it A Modern Point of View on Anomalies,}
\href{https://doi.org/10.1002/prop.201910012}{Fortsch. Phys. \textbf{67} (2019) 1910012}
[\href{https://doi.org/10.48550/arXiv.1903.02828}{arXiv:1903.02828 [hep-th]}].

\bibitem{discovery}
S.~L.~Adler,
{\it Axial vector vertex in spinor electrodynamics,}
\href{https://doi.org/10.1103/PhysRev.177.2426}{Phys. Rev. \textbf{177} (1969) 2426}. 
\\
J.~S.~Bell and R.~Jackiw,
{\it A PCAC puzzle: $\pi^0 \to \gamma \gamma$ in the $\sigma$ model,}
\href{https://doi.org/10.1007/BF02823296}{Nuovo Cim. A \textbf{60} (1969) 47}.

\bibitem{AS}
N.~K.~Nielsen and B.~Schroer,
{\it Axial Anomaly and Atiyah-Singer Theorem,}
\href{https://doi.org/10.1016/0550-3213(77)90453-9}{Nucl. Phys. B \textbf{127} (1977) 493}.
\\
N.~K.~Nielsen, H.~R\"omer and B.~Schroer,
{\it Classical Anomalies and Local Version of the Atiyah-Singer Theorem,}
\href{https://doi.org/10.1016/0370-2693(77)90410-5}{Phys. Lett. B \textbf{70} (1977) 445}.
\\
K.~Fujikawa,
{\it Path Integral Measure for Gauge Invariant Fermion Theories,}
\href{https://doi.org/10.1103/PhysRevLett.42.1195}{Phys. Rev. Lett. \textbf{42} (1979) 1195}. 
\\
K.~Fujikawa,
{\it Path Integral for Gauge Theories with Fermions,}
\href{https://doi.org/10.1103/PhysRevD.21.2848}{Phys. Rev. D \textbf{21} (1980) 2848}.

\end{thebibliography}
\end{document}